\documentclass{IEEEtran}

\usepackage{cite}
\usepackage{graphicx}
\usepackage{subfigure}

\usepackage{threeparttable}
\usepackage{amsthm}
\usepackage{mathrsfs}
\usepackage{amssymb}
\usepackage{cite}

\ifCLASSINFOpdf
\else
   \usepackage[dvips]{graphicx}
   \graphicspath{{../eps/}}
   \DeclareGraphicsExtensions{.eps}
\fi

\usepackage[cmex10]{amsmath}
\usepackage{empheq}

\usepackage{mdwmath}
\usepackage{mdwtab}
\usepackage{amsfonts}
\usepackage{algorithm}
\usepackage{algorithmic}

\renewcommand{\algorithmicrequire}{\textbf{Input:}}
\renewcommand{\algorithmicensure}{\textbf{Output:}}

\usepackage{algorithmic}

\makeatletter
\newcommand{\removelatexerror}{\let\@latex@error\@gobble}
\makeatother

\begin{document}
%
\title{Federated Meta Learning Enhanced Acoustic Radio Cooperative Framework for Ocean of Things Underwater Acoustic Communications}


 

\author{  Hao Zhao, Fei Ji, Quansheng Guan, Qiang Li, Shuai Wang,  Hefeng Dong and Miaowen Wen}
\maketitle

\begin{abstract}
Sixth-generation wireless communication (6G) will be an integrated architecture of "space, air, ground and sea". One of the most difficult part of this architecture is the underwater information acquisition which need to transmitt information cross the interface between water and air.
In this senario, ocean of things (OoT) will play an important role, because it can serve as a hub connecting  Internet of things (IoT) and Internet of underwater things (IoUT). OoT device not only can collect data through underwater methods, but also can utilize radio frequence over the air.  For underwater communications, underwater acoustic communications (UWA COMMs) is the most effective way for OoT devices to exchange information, but it is always tormented by doppler shift and synchronization errors. In this paper,  in order to overcome UWA tough conditions, a  deep neural networks based receiver for underwater acoustic chirp communication, called C-DNN, is proposed.  Moreover,  to improve  the performance  of  DL-model and solve the problem of model generalization, we also proposed a novel federated meta learning (FML) enhanced acoustic radio cooperative (ARC) framework, dubbed ARC/FML, to do transfer. Particularly, tractable expressions are derived for the convergence rate of FML in a wireless setting, accounting for effects from both scheduling ratio, local epoch and  the data amount on a single node.
From our analysis and simulation results, it is shown that, the proposed C-DNN can provide a better BER performance and lower complexity  than  classical matched filter (MF) in underwater acoustic communications scenario. The ARC/FML framework  has good convergence under a variety of channels than federated learning (FL). In summary, the proposed  ARC/FML for OoT is a promising scheme for information exchange across water and air.
\end{abstract}



%
\IEEEpeerreviewmaketitle

\section{Introduction}

With the continuous developmenting of the blue economy, the communications for marine information gathering, transmission and fusion will become more and more important\cite{uwanet}.
Moreover, sixth-generation (6G) agenda aims to  connect the  whole world together, which needs to ensure worldwide connectivity. The  demand  of  commnunications will expand from space to air, ground, and sea environment in this era, dramatically\cite{6Gtvt}~\cite{6Gnetwork}. 
Hence, underwater information  acquisition is becoming increasingly important. Underwater acoustic communication (UWA COMMs)  as the most effective way of underwater information transmission  has two challenges to face.
\begin{figure}[t!]
	\centering
	\includegraphics[width=9cm]{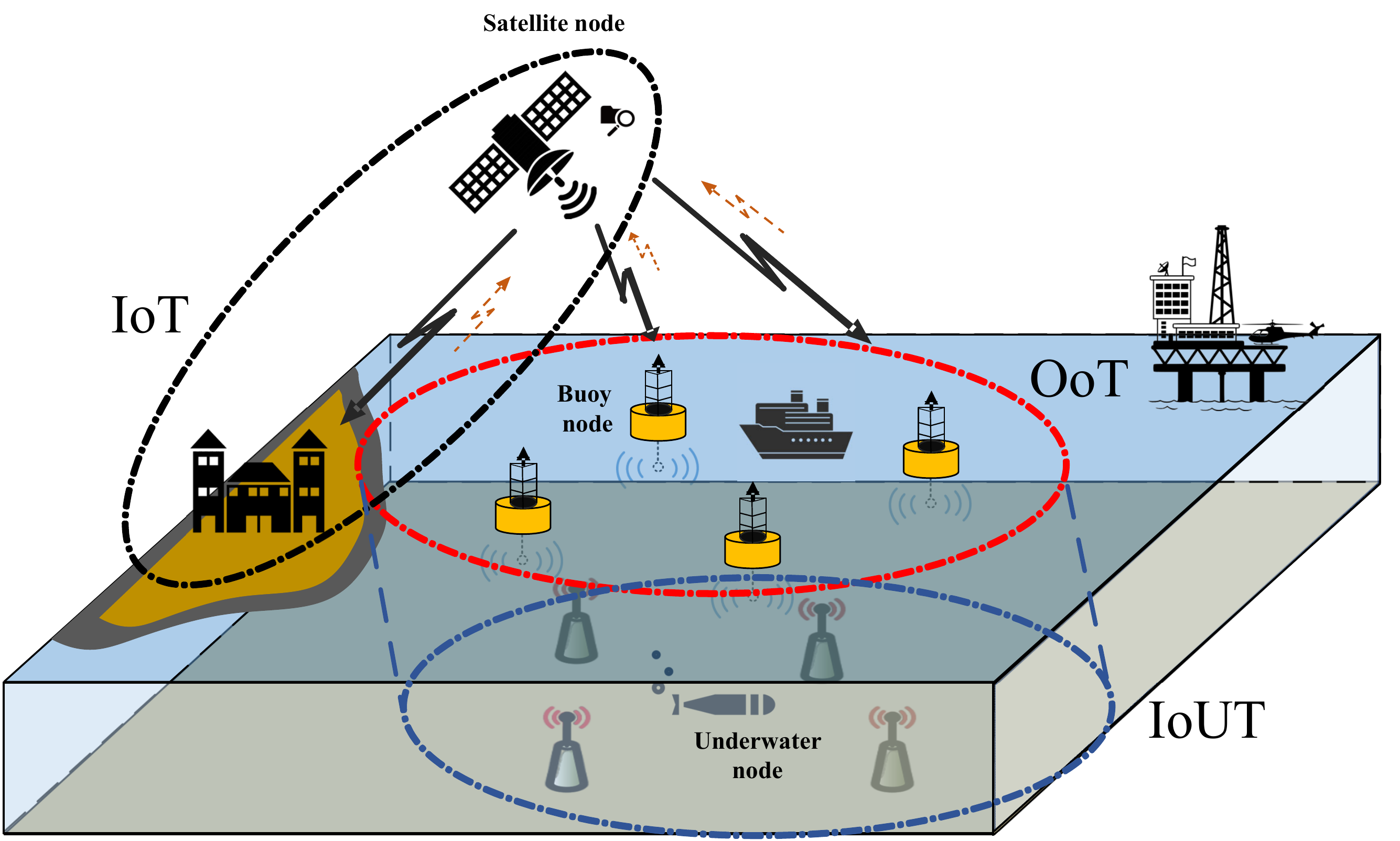}\\
	\caption{Distributed OoT devices link IoT devices and IoUT devices.}\label{RelayIllustration}
\end{figure}

One challenge  is   underwater information need cross the water-air interface.
Fortunately, using buoy node to exchange seabed observation information with satellite control information is a very typical and extremely important application for ocean observation \cite{situOB}. 
The emerging ocean of things (OoT) based on low-cost floating devices \cite{ootJ} \cite{ootoceans}, will provide a feasible way for water and air information interaction.
Therefore, OoT will be the hub of information interaction, linking  IoT, which is for wireless devices and IoUT which is mainly focusing on underwater equipment\cite{IoUT}.

Another challenge is that UWA COMMs need overcome many obstacles, such as strong noise interference,  multipath effects, large scale Doppler effects and synchronization error\cite{Chitre2008Underwater}.
The UWA COMMs can be divided into two research fields, high rate communication system and low rate robust communication system.
Especially, the robust UWA COMMs play an important role in many  underwater scenarios, such as control signaling transmission for unmanned underwater work system and information interaction in a high-noise environment.
Recently, deep learning   has shown amazing results in solving underwater acoustic signal recovery than classical signal processing method. However, the fatal problem is that devices are distributed and may have insufficient data in single node.
Hence, there are two important problems that can't be ignored.  First, data is separated which lead to the marginalization and discretization of data acquisition. Second, single device may have insufficient data. 
The emerging federated learning (FL) can  train deep learning (DL) models in distributed systems which is a good solution\cite{FLgen}.

Motivated by above mentioned,   we explore the power of deep learning and exploit the cooperation of acoustic and radio links to use distributed data to achieve robust UWA COMMs and utilize distributed data. The surface relay buoy transmission system can utilize the cooperation of acoustic and radio. First, the system can realize information interaction with the subsea equipments through DL based UWA COMMs. Second, the sea surface relay buoy can do federated learning to share DL model parameters via radio frequence in order to improve single node performance. In this paper, the main contribution can be divided into three parts.

 

%

$\bullet$  We proposed an acoustic radio cooperative (ARC) training framework for deep learning based Ocean of Things, which can be used to  DL-model training for surface equipment.

$\bullet$  To analyze the ARC performance, we take stability UWA COMMs for OoT device as an example.  We propose a novel DL-based  chirp communications receiver  apply it   over underwater acoustic channels, which can against doppler shift and symbol time offset. The bit error rate can be increased by an order of magnitude.


$\bullet$  To utilize the distributed data from multiple  buoy nodes,   we proposed an ARC enhanced federated meta learning (FML) based algorithm to train the DL-receiver in the context of random scheduling wireless networks, dubbed  ARC/FML, which can achieve distributed transfer learning  to adapt to a new dataset.  Besides, we analysis the converges of FML with wireless communication. For any convergence target gap $\epsilon$, the FML algorithm  can acheive an gap after $T_z$ rounds of communications.

The remainder of this paper is organized as follows. In section II, provides a brief survey on UWA COMMs, DL in physical layer and federated learning in wireless networks. Then, the system model is introduced in section III and the convergence  of Federated Meta Learning in Wireless  is analysed on section IV. In section V, the  dataset is explained.In section VI  simulation results are demonstrated. In last section, a conclusion is provided.

\section{State of Art}

\textbf{UWA COMMs:} UWA characteristics are now known the disadvantages of severe transmission loss, time-varying multi-path propagation, severe Doppler spread, limited and distance-dependent bandwidth, and high propagation delay\cite{WENM}. These features  will change as the communication scenario changes. Therefore, researchers usually divide underwater acoustic communication into two kinds according to application requirement \cite{Huang}. One is high-data-rate underwater acoustic communication for short- and medium-range communications. In this direction,in order to further improve the communication reliability, it is necessary to better overcome the complex multi-path fading of underwater acoustic channel. The joint equalization decoding method, which combines channel equalization and channel decoding, is a relatively advanced channel compensation technology at present and has high practical value and application prospect\cite{TURBO}. Moreover, in order to meet the needs of more diversified applications with high data rate, new efficient modulation mode and multiple input multiple output technology are introduced into the medium and short range underwater acoustic communication, which significantly improves the rate of underwater acoustic communication and becomes a new research hotspot\cite{MIMOuwa}.
The other is low-data-rate underwater acoustic communication for long-range communications. Many modulation techniques which have robust performance, such as frequence shift keying, chirp modulation and spread modulation, have been used for rapid timevarying channels.
However, the most difficult aspect of UWA COMMs is lack of a accurate channel models
the tractable mathematical descriptions of the underwater acoustic channel are elusive, because the  signal propagation is very complicated\cite{NarimanNeural}.
Hence, researchers pay more attention to data based deep learning method expecting to solve many problems that traditional methods cannot.
With the improving computational resources  and the quantity of data,  deep learning  brings a new era for communication system that many novel system architecture and algorithm are designed  \cite{o2017introduction}.

\textbf{DL Based COMMs:} 
DL has been applied successfully in receiver design, channel estimation  and signal detection over wireless channels~\cite{qin2019deep}. Unlike conventional receivers, DL can handle wireless channels in an end-to-end manner. It is widely acknowledged that the well-trained DL-receiver can not only reduce the receiver complexity  \cite{TwoApplications}, but also achieve perfect demodulation under unknown channels \cite{NarimanNeural}. 
The design of DL-receivers can be generally classified into  two categories. One is the data driven method, such as the classical FC-DNN\cite{ye2017power}, which takes into account the characteristics of data and the ability  of neural networks, aiming to achieve global optimality. Whereas, most existing works based on the data-driven method consider the communication system as a black box. The other category is the model driven method, such as the well-known ComNet~\cite{Comnet}, which  combines DL and expert knowledge. 
In especial, some methods based on deep learning are also gradually being used in underwater acoustic communication. 

\textbf{FL:} 
FL  come into fashion because it can decouple the data acquisition and compute at the central unit.  
An analytical model is developed to characterize the performance of FL in wireless networks\cite{FLtony}. 
Moreover, to ensure the DL based communications can  work at a new environment.  Meta-learning, which can train the network by alternating inner-task and across-task updates  with a small number of labeled data  to improve the transfer efficiency, has been used in wireless communications.  
Therefore, many researchers pay attention to federated  meta learning.
However, among many applications in federated learning based wireless network, deep learning (DL) based applications are actively and widely discussed on image classification task, such as MNIST, rather than  DL based physical layer cases. 
Recently, in the ocean of things, with the increasing computational capacity of device, such as buoys, unmanned ships and offshore platform,  as well as the increasing concerns about sharing private data. There is a precedent that the IoUT device  can realize federated learning (FL) computation \cite{UWAiot}.

\section{Federated Meta Learning Based Framework}

In this section, we introduce the  federated meta learning in wireless algrithom.
\begin{figure}[t]
	\centering
	\includegraphics[width=8cm]{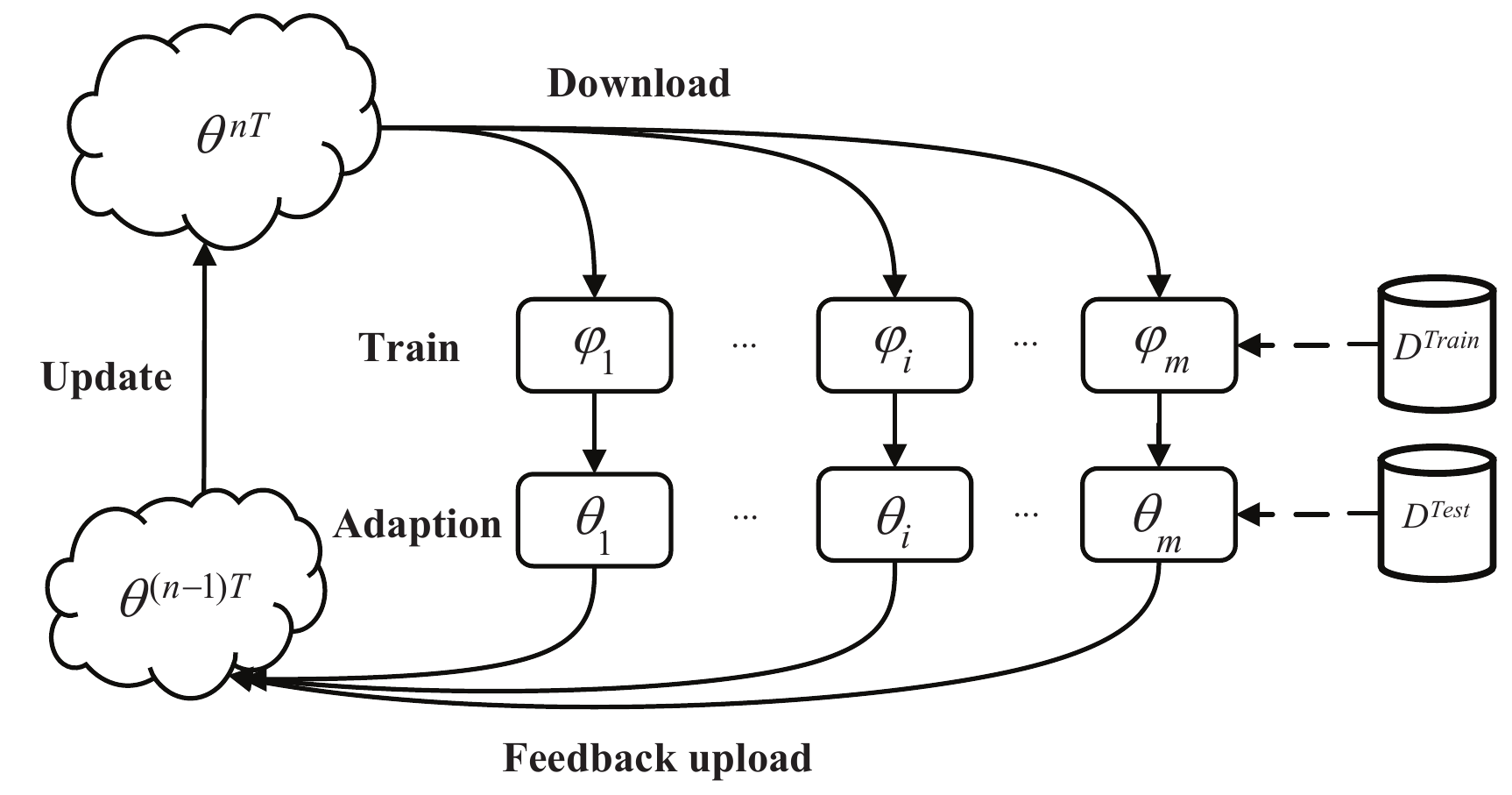}\\ 
	\caption{Federated meta learning framework.}\label{RelayIllustration}
\end{figure}
For DL based applications, an important limitation of the approach is that training should be generally carried out from scratch for each new dataset. Aiming at improve the generalization of DL-receiver, we proposed FML based framework in random scheduling.
Squentially updating the network parameters from all edge nodes then do global aggregation which is very efficient than transimitted all data loacated on distributed node to the centre node. In many communication scenarios, the channel sources  are precious.  Hence, sheeduling police is used to allocate the limited channels to users. In \cite{FLtony}, the federated learning in wireless network  with three scheduling policies have been studied and the convergences are analyzed.
However, existing federated learning focus on a sufficient dataset and is hard to adapt to  new UWA environment when labeled data is rarely.
The issue of robust learning is highly  concerned for machine learning for communications. Many approaches have been proposed, in particular the approach of coarse offline learning using data sets and fine learning in an online fashion have been proposed. 
The goal of federated meta learning fits in well with this, which is to train models performing one or a few steps
of gradient descent on their local dataset and  such that it can solve new learning tasks using only a small number of training samples\cite{DTLaibo}.
Hence, in this paper, we take federated meta learning as the framework for distributed model with random scheduling. In each communication rounds, the center node will uniformly pick $N$ users out of $K$ users and $G=\frac{N}{K}$ is the available channel ratio.  
Essentially, federated meta learning is used to do transfer learning in order to improve the generalization of DL-receiver. 
Here, we focus on applying FML in wireless networks.  The procedure is description in algorithm 1. The algorithm can be divided into two parts.

$\bullet$ At edge node $i$, it first update using the training data $D_i^{training}$ stored on the device. 
For MAML algorithm, given that the model parameters of $i$ buoy node, the node can update its  parameters by one step learning  according gradient descent based on $D_i^{train}$,
\begin{equation}
\phi_i(\theta)=\theta -\alpha \nabla_\theta L(\theta, D_i^{train}),
\end{equation}
where $\alpha$ is the learning rate and then evaluates the loss $L(\phi_i, D_i^{train})$.
Then, locally update $\theta_i$ using testing data $D_i^{test}$: 
\begin{equation}
\theta_i^{t+1}= \theta_i^t-\beta \nabla_\theta L(\phi_i(\theta), D_i^{test}).
\end{equation}
After that, if the node is choosen by AP, it will send  $\theta_i^{t+1}$ to the AP.  The framework can be seen in Fig.3.

$\bullet$ At the AP side,  it selects part of BNs for collect parameters in order to the global aggregation. 
During a communication round $t$,  the parameters can be uploaded and updated successfully which  needs to meet two conditions, simultaneously.   One is that the node $k$  should be selected and the other is that the  transmitted data should  be decoded without error.  In this respect, we use indicator function $u_{i}^t\in \{0,1\}$ to mark the node $i$ which will be used in  federated learning process, 
which indicate   whether the node  $i$  be chosen at  $t$-th round. Hence, the AP performs
\begin{equation}
\theta^{t+1} =\sum _{ i \in \mathbb{S}}w_i \theta_i^{t+1} u_i,
\end{equation}
where  $w_i$  can be calculated by the local  data size according by  $w_i= \frac{|J_i|}{\sum_{i \in \mathbb{S} }||J_i||}$.
Then, the new global paprameters $\theta^{t+1}$ will be broadcast to all nodes in a reliable way.

We first assume the distributed buoy  $i \in \mathbb{S}$ and $J_i$ represents its local dataset ${(x_i^{1}, y_i^{1}),...,(x_i^{j}, y_i^{j}),...,(x_i^{J_{i}}, y_i^{J_{i}})}$, where $|J_i|$ means dataset and $(x_i, y_i)$ is the sample of this dataset. $x_i$ is the input of DL-based network which is the receive signal and  $y_i$ is the output of DL-based network. The distribution of the dataset is unknown. The loss function can be defined as $l(\theta, (x_i, y_i)) :\mathcal{X} \times \mathcal{Y} \rightarrow \mathbb{R} $. 
The  receiver located on buoys can be trained by optimizing the loss function as
\begin{equation}\label{eq-loss-each}
L_i(\theta, J_j) \triangleq \frac{1}{|J_j|}\sum_{\substack(x_i^j, y_i^j)\in |J_j|}{{{\left| {{x_i^j} -  {\hat x_i^j}} \right|}^2}}.
\end{equation}
which can be simply denoted as $L_i(\theta)$.


\begin{algorithm}[!t]
	\caption{Federated Meta Learning Based on Random Scheduling}
	\label{alg:example}
	\begin{algorithmic}[1] 
		\REQUIRE  {\bf{Data set $\{D_i\}_{i=1}^{i=K}$} at each BN} \\ 
		\FOR {$ t=1:T $}
		\FOR {each UE $k \in \{ 1,2, \ldots ,K\}$ in parallel}
		\STATE {Initialize $w_i^t = {w^t}$}
		\FOR {$t_{ local}=1$ to $T_0$  }	
		\STATE {Sample $i \in D_i$ uniformly at random, and update the local parameter using $D_{train}$}
		\STATE {$\phi_i^t = \theta_i^t - \alpha (\nabla_\theta {L}(\theta_i^t), D_{train})$}
		\STATE {obtain $ \theta_i^{t+1}$ based  on}
		\STATE {$\theta_{i+1}^t = \theta_i^t - \beta (\nabla_\theta \phi_i^t , D_{test})$}
		\ENDFOR
		
		\STATE {Send parameter $w_i^t$ to the AP }
		\ENDFOR
			\STATE{The AP collects all the parameters $\{ \theta_i^t\} _{i = 1}^K$, and updates ${\theta^{t + 1}} = \frac{1}{n}\sum\limits_{i = 1}^K {{w_i}\theta_i^t}u_i $ } 
	
		\ENDFOR  
		\ENSURE $\theta^T$      
	\end{algorithmic}
\end{algorithm}

\section{UWA Chirp Communications Cases}
With the advance of deep learning, it  has a very strong application in the physical layer.
Hence, we take DL-based communications system as an example.
In order to achieve the purpose of stable communication, each node selects chirp communications for UWA COMMs, because of its characteristic of  anti-noise and  robustness to Doppler.
The UWA stable communications adopt a pair of chirp signals in frequency band $[f_1 - f_2]$ Hz to transmit information which can be expressed as
\begin{figure}[t]
	\centering
	\includegraphics[width=9cm]{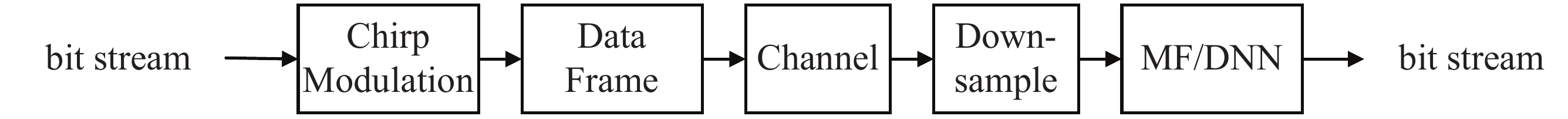}\\  
	\caption{Chirp communication system framework.}\label{RelayIllustration}
\end{figure}
\begin{equation}\label{eq-III.1}
s_1(t)= cos(2 \pi f_1 t+\mu t^2/2 +\phi_0 ), 0 \leq t\leq T,
\end{equation} 
\begin{equation}\label{eq-III.1}
s_2(t)= cos(2 \pi f_2 t-\mu t^2/2 +\phi_0 ), 0 \leq t\leq T,
\end{equation} 
where $s_1$ means up chirp and $s_2$ is down chirp where $T$ is symbol duration, and $\phi_0$ is arbitrary initial phase which is assumed to be zero without loss of generality, respectively.
Moreover, the chirp signal is characterized by its start frequency $f_1$, end frequency $f_2$, and time duration $T$ as
\begin{equation}\label{eq-III.1}
\mu =\frac{|f_2-f_1|}{T}=\frac{B}{T},
\end{equation} 
where $B$ is the bandwidth of chirp signals. Usually, the
chirp signal is defined as up-chirps with $\mu > 0$  and
down-chirps with $\mu < 0$.
After  framed, the signal is transmitted over the channel which can be expressed as 

\begin{equation}\label{channelintergral}
y(t)= \int_{-\infty}^{+\infty}h(\tau,t)s(t-\tau)d\tau +n(t),
\end{equation} 
where $ h(t) $ means UWA channel impulse response.

\subsection{Matched Filter based Receiver}
A classical method to implement the matched filter (correlator) receiver is convolution. Received signal $y(t)$ is convolved with time-reversed versions of $s_1(t)$ and $s_2(t)$ to generate estimator $c_1$ and $c_2$, respectively. The receiver calculation process  is as follow,

\begin{equation}\label{eq-III.1}
c_{i}= \int y(\tau)s_{i}(T-\tau)d \tau, ~~i \in {(1,2)},
\end{equation}

If $c_1 \leq c_2$, the receiver estimates that $s_1$ is transmitted. Otherwise, $s_2$ is transmitted.
However, to achieve the optimal detection, the matched filter needs to satisfy three conditions. 

$\bullet$ 	Integral interval synchronization should be satisfied. That is to say, the received signals $y(t)$  require precise synchronization. 

$\bullet$ The noise $n(t)$ should be Gaussian noise. 

$\bullet$ 	The received signal $y(t)$ must not be affected by Doppler shift.

Unfortunately, the abovementioned assumptions are impractical in UWA COMMs, because of the large delay, non-Gaussian noise and  doppler shift charactersitc. Hence, we proposed a novel DL-receiver solve those problems.
%
%

\subsection{DL based Receiver}
We first introduce a specific  DL based  receiver, which  depends on  neural network  architecture, called C-DNN, which  can be employed for
detection uses several fully connected NN layers. Hence, we assume that the DL-receiver can be denoted as $f_\theta$ with parameters $\theta$ $\in \mathbb{R}^d$.
In this paper, primitively, let us consider the simplest four-layer fully-connected
neural network with  one input layer, two hidden
layers, and one output layer. Denote $x_i$ and $y_i$
as the estimited data and the network input of node $i$,
respectively.
Therefore, the receiver $f(\theta)$ which is a cascade of nonlinear
transformation of input data can be expressed as
\begin{equation}
\hat x_i= f(y,\theta)=f_{sigmoid}^{(L-1)}(f_{Relu}^{(L-2)}(...f_{Relu}^{(1)}(y_i))),
\end{equation}
where layer $l$ contains a total
of $N_l$ neurons and each neuron in the layer $l$ is connected to all
neurons in the next layer $(l+1)$ through the connection weights
matrix.
The output of hidden layer are activated by 
\begin{equation}
f_{Relu}(x_i) = \max (0,x_i),
\end{equation}
which is a non-linear function to provide a normalized output
and keeps the output within the interval $[0, +\infty]$.
The output of the final layer is the estimated bit.
The output layer only consist one neuron, estimating the binary bits to be detected.
Considering sigmoid activation function is applied to the output layer, which limits the output within the interval  $[0, 1]$. Thus, the  task can be activated by 
\begin{equation}
f_{sigmoid}(x_i) = \frac{1}{{1 + {e^{ x_i}}}}.
\end{equation}
The input to the first layer
is the received signal  $y_i(t)$ or the sampled feature factor $y_i[k]$, which
is selectively choosen from the observed signal through
preprocessing.
This dataset is then used to train a DL-based receiver that
estimate the received signal $y_i[k]$ to one of the transmission symbols in $[s_1, s_2]$. 
\begin{table}[b]
	\centering
	\renewcommand\arraystretch{1.2}
	\renewcommand\tabcolsep{3.0pt}
	\setlength{\abovecaptionskip}{0.cm}
	\caption{Notation Summary.}
	\begin{tabular}{p{2cm}p{6cm}}
		\hline
		Notations & Definition \\
		\hline
		$T_0$ &	number of local update steps\\
		$v^t_{[n]}$	&	parameters for global aggregation at each iteration within the interval $[(n-1)T_0,nT_0]$\\
		$\theta_i^t$; $\theta^t$ &  local parameters; weighted average of local parameters which is synchronized with $v^t_{[n]}$ \\
		$\alpha$; $\beta$ &  learning rate during meta training (local adaptation);	 learning rate during meta training (update initialization)\\
		$L_i$; $G_i$; $G$ & objective function\\
		$D^{train}$; $D^{test}$ & dataset for training and dataset for testing\\
		
		\hline  
	\end{tabular}\\
\end{table}

\section{Convergence Analysis of Federated Meta Learning}

In this section, we focus on the convergence of the federated meta-learning method.  We first define $G_i(\theta) = L_i(\phi_i(\theta))$ and  $G(\theta) = \sum_{ i \in \mathbb{S}} w_i G_i(\theta)$. For  simplicity, we assume $T=NT_0$ and do four assumptions for all objective functions.

\textbf{Assumption 1.} Each $L(\theta)$ is $\mu$-strong convex, i.e., for all $\theta$, $\theta'\in \mathbb{R}^n$, 
\begin{equation}
||\nabla L_i(\theta)-\nabla L_i(\theta')|| \geq \mu ||\theta- \theta' ||.
\end{equation}

\textbf{Assumption 2.} Each $L(\theta)$ is $H$-smooth, i.e., for all $\theta$, $\theta'\in \mathbb{R}^n$, 
\begin{equation}
||\nabla L_i(\theta)-\nabla L_i(\theta')|| \leq H ||\theta- \theta' ||.
\end{equation}
and there exists constant $B$ such that for all $\theta'\in \mathbb{R}^n$,
\begin{equation}
||\nabla L_i(\theta)|| \leq B.
\end{equation}

\textbf{Assumption 3.} The hessian of each  $L(\theta)$ is $\rho$-Lipschitz, i.e., for all $\theta$, $\theta'\in \mathbb{R}^n$, 
\begin{equation}
||\nabla^2 L_i(\theta)-\nabla^2 L_w(\theta)||\leq\ \rho||\theta- \theta' ||.
\end{equation}

\textbf{Assumption 4.} There exists  $\delta$ and $\sigma$ such that for all $\theta \in \mathbb{R}^n$, 
\begin{equation}
||\nabla L_i(\theta)-\nabla L_w(\theta)||\leq\delta_i  
\end{equation}
\begin{equation}
||\nabla^2 L_i(\theta)-\nabla^2 L_w(\theta)||\leq\sigma_i
\end{equation}

Assumption 1 and 2 are standard in standard and hold in many deep learning algorithm. 
Assumption 3 illustrates that the local loss function is second-order smooth which is  possible  to analyse local meta learning loss function.
Assumption 4 describes the node similarity, whose gap between  an implementation and the sample average can be measured by $||\nabla L_i(\theta)-\nabla L_w(\theta)||$.

Next, to analysis the convergency characteristic of FML based on random scheduling, we first characteristic the global loss function $G(\theta)$. 
Here, we first show that $G(\theta)$ is $\mu''$-strongly convex and $H''$-smooth.

\textbf{Lemma  1.}
Suppose assumption 1-3 hold.
$G(\theta)$ is $\mu''$-strongly convex and $H''$-smooth, where $\mu''=N \mu'$, $ H''=N H'$ , $\mu' =\mu (1 - \alpha H)^2- \alpha \rho B $  and $ H'= H(1-\alpha \mu)^2+ \alpha \rho B$.

Lamma 1 tell us that the total loss function $G(\theta)$ is also  a convex function as $L(\theta)$

\emph{\textbf{Proof:}}
To establish  the smooth, we should show $||\nabla G(\theta)- \nabla G(\theta')|| \leq H'' ||\theta-\theta'||.$
By the definition of $H$-smooth,   we have
\begin{equation}
\begin{aligned}
G_i(\theta) \leq G_i(\theta') +\nabla G_i(\theta')(\theta - \theta')+\frac{H'}{2}||\theta-\theta'||^2,
\end{aligned}
\end{equation} 
which is equivalent to
\begin{equation}
\begin{aligned}
	||\nabla G_i(\theta)- \nabla G_i(\theta')|| \leq H'||\theta-\theta'||,
\end{aligned}
\end{equation} 
where $i \in \mathbb{S}$.
Because  $G(\theta) = \sum_{ i \in \mathbb{S}} w_i G_i(\theta)$,  by summing we can get
\begin{equation}
\begin{aligned}
\sum_{ i \in \mathbb{S}}  G_i(\theta) &\leq   \sum_{ i \in \mathbb{S}} G_i(\theta') +  \sum_{ i \in \mathbb{S}} \nabla G_i(\theta')(\theta - \theta')\\&+ \frac {\sum_{i=1}^{N} H'}{2}||x-y||^2.
\end{aligned}
\end{equation}
That is to say,
\begin{equation}
\begin{aligned}
G(\theta) \leq G(\theta') +\nabla G(\theta')(\theta - \theta')+\frac{N H'}{2}||\theta - \theta'||^2,
\end{aligned}
\end{equation}
which is equivalent to
\begin{equation}
\begin{aligned}
||\nabla G(\theta)- \nabla G(\theta')|| \leq H''||\theta-\theta'||,
\end{aligned}
\end{equation} 
where $H''= NH'$.

In the same way, we can  establish  the convex, 
\begin{equation}
\begin{aligned}
 ||\nabla G(\theta)- \nabla G(\theta')|| \geq N\mu'||\theta-\theta'|| .
\end{aligned}
\end{equation}
From abovementioned, we can have
\begin{equation}
N\mu'||\theta-\theta'|| \leq ||\nabla G(\theta)- \nabla G(\theta')|| \leq NH'||\theta-\theta'||.
\end{equation} 
Thereby we complete the proof.

Next, we analysis the influence of the similarity between local learning tasks.
Based on Lemma 1, we can get the convergence target gap  of the FML based method.

\textbf{Theorem  1.} For any convergence target gap $\epsilon$, the FML algorithm  can acheive  the gap after $T_z$ rounds of communications, i.e., 
\begin{equation}
\mathbb{E}[G(\theta^*) - G(\theta^T)] \leq \epsilon
\end{equation}
if  $T_z$  satisfies the following 
\begin{equation}\label{impans}
T_z \geq \frac{log(\frac{1}{n}(\epsilon + K m(T)))}{log(\xi)},
\end{equation}
where  $K = \frac{ \mu'' } {1-\xi^{T_0}}$,  $\xi =1-2H'' \beta(1+\frac{\mu''\beta}{2})$ and $m(T)= \alpha' T-\frac{\alpha'}{\beta H'}[1-(1-\beta H')^{T}]$.

The proof is as follows. From formula (\ref{impans}), we can find that the term $K$ is influenced by  the diffence of meta task and the
mutiple local update using the function $m(T_0)$.
According to the $m(T_0)$, we can find that the local step $T_0$ impact the convergence time $T_z$. With the increasing of $T_0$, the 
$T_z$ decrease. Hence, we can adjust the $T_0$ to balance transmission cost and local calculation cost.

\emph{\textbf{Proof:}}
Considering the same ways of \cite{AdaptiveFL},  \cite{realtimeFL}, we define virtual sequence $v^t_{[n]}$ for global aggregation at each iteration  for $t \in [(n-1)T_0,nT_0]$. the internal $[(n-1)T_0,nT_0]$ is regarded as $[n]$. In general, we have
\begin{equation}
v^{t+1}_{[n]} = v^t_{[n]} -\beta \nabla G(v^t_{[n]} ),
\end{equation}
where $v^t_{[n]}$is assumed to be "synchronized" with $\theta^t$ at the beginning of interval $[n]$ ,i. e,  $v^t_{[n]}  = \theta^{(n-1)T_0}$, where $\theta^{(n-1)T_0}$ is the globle averaging model parameters   $\theta_i^{(n-1)T_0}$. 

To show the convergence, we first analyze the gap between  $v^t_{[n]}$ and $\theta^t$,
\begin{equation}
\begin{aligned}
&||\theta^{t+1}_i - v^{t+1}_{[n]}||\\
& =  || \theta^{t}_i- \beta \nabla G_i(\theta_i^t) -v_{[n]}^t +  \beta \nabla G(v_{[n]}^t)||\\
&\leq || \theta^{t}_i-v_{[n]}^t|| + \beta || \nabla G(v_{[n]}^t)-  \nabla G_i(\theta_i^t)||\\
&\leq || \theta^{t}_i-v_{[n]}^t|| +   (\beta ||\nabla G_i(\theta_i^t) - \nabla G_i(v_{[n]}^t) || \\
&~~~+ \beta || \nabla G_i(v_{[n]}^t)- \beta || \nabla G(v_{[n]}^t)||)\\
&\leq (1+\beta H') ||\theta^{t}_i - v^{t}_{[n]}||\\&+ \beta [\delta_i +  \alpha C (H \delta_i+B\sigma_i +\tau)]
\end{aligned}
\end{equation}
where the upper bound of $||\theta^{t}_i - v^{t}_{[n]}  ||\leq \delta_i +  \alpha C (H \delta_i+B\sigma_i +\tau) $ can be found from \cite{realtimeFL}.

Next, we denote $g(x) \doteq \frac{\delta_i +  \alpha C (H \delta_i+B\sigma_i +\tau)}{H'}[(1+\beta H')^x-1]$.  and we can have $||\theta^{t}_i - v^{t}_{[n]}|| \leq g(t-(n-1)T_0)$. According this, we can get
\begin{equation}
\begin{aligned}
&||\theta^{t+1} - v^{t+1}_{[n]}||\\
&= ||\sum w_i \theta^{t+1}_i u_i^t - v^{t+1}_{[n]}||\\
&=|| \theta^{t}- \sum w_i \nabla G_i(\theta_i^t) u_i^t - v^{t}_{[n]}+ \beta \nabla G(v_{[n]}^t) ||\\
&\geq || \theta^{t}-v_{[n]}^t|| - \beta || \sum w_i  ( \nabla G_i(\theta_i^t) - \nabla G_i(v_{[n]}^t)) ~u_i^t||\\
&\geq  || \theta^{t}-v_{[n]}^t|| - \beta H' \sum_{i \in \mathbb{S}} w_i||\theta^{t}_i-v^{t}_{[n]}|| u_i^t\\
&\geq  || \theta^{t}-v_{[n]}^t||- \beta H' \sum_{i \in \mathbb{S}} w_i||\theta^{t}_i-v^{t}_{[n]}|| \\
&\geq  || \theta^{t}-v_{[n]}^t|| - \beta H' \sum_{i \in \mathbb{S}} w_ig(t-(n-1)T_0) \\
&= || \theta^{t}-v_{[n]}^t||+  \alpha'[1-(1+\beta H')^{t-(n-1)T_0}]
\end{aligned}
\end{equation}
where $\alpha' = \beta [\delta +\alpha C(H \delta +B  \sigma +\tau)]$. Iteratively, we have 
\begin{equation}
\begin{aligned}
&||\theta^{t} - v^{t}_{[n]}|| \geq \sum_{j=1}^{j=t-(n-1)T_0}{\alpha'[1-(1+\beta H')^{j}]}\\
&= \alpha' [(t-(n-1)T_0]-\frac{\alpha'}{\beta H'}[1-(1-\beta H')^{(t-(n-1)T_0)}]\\
&\doteq m(t-(n-1)T_0)
\end{aligned}
\end{equation}
Then, we analyse the gap between virtual sequence $v^t_{[n]}$ and  $v^{t+1}_{[n]}$ within the interval $[n]$ at $t \in [(n-1)T_0,nT_0]$. Because $G(.)$ is $H''$-smooth, we have
\begin{equation} \label{com1}
\begin{aligned}
&  G(v^{t}_{[n]}) -G(v^{t+1}_{[n]})\\
&\leq \nabla G(v^t_{[n]})(v^{t}_{[n]}- v^{t+1}_{[n]})  + \frac{H''}{2}||v^{t}_{[n]}-v^{t+1}_{[n]}||^2\\
&\leq \beta(1+\frac{H''\beta}{2})||\nabla G(v^{t}_{[n]}) ||^2
\end{aligned}
\end{equation}

Assumption 3 told us that $G(.)$ is $H''$-smooth and $\theta^*$ is the minimum point, we have
\begin{equation}\label{3232}
\begin{aligned}
G(\theta^*)&= \min_{\theta}  G(\theta) \\
&\leq  \min_{\theta} [G(v^{t}_{[n]}) +\nabla G(\theta')(\theta - v^{t}_{[n]})+\frac{H''}{2}||\theta-v^{t}_{[n]}||^2]\\
&= G(v^{t}_{[n]}) + \min_{||y||=1}  \min_{t\geq 0}[\nabla G(v^{t}_{[n]})yt+\frac{H''}{2}t^2 ]\\
&= G(v^{t}_{[n]}) + \min_{||y||=1} [-\frac{(\nabla G(v^{t}_{[n]})y)^2}{2H''}]\\
&= G(v^{t}_{[n]}) - \frac{||\nabla G(v^{t}_{[n]})||^2}{2H''},\\
\end{aligned}
\end{equation}
where $t=\theta-v^{t}_{[n]}$.
Therefore, we can get
\begin{equation}\label{com2}
\begin{aligned}
\frac{1}{2H'' }||\nabla G(v^{t}_{[n]})||^2 \leq G(v^{t}_{[n]}) - G(\theta^*) 
\end{aligned}
\end{equation}

Combined formula (\ref{com1}) and (\ref{com2}), we can have
\begin{equation} \label{oooooottt}
\begin{aligned}
G(v^{t}_{[n]})- G(v^{t+1}_{[n]}) \leq 2H'' \beta(1+\frac{H''\beta}{2}) [ G(v^{t}_{[n]}) - G(\theta^*) ]
\end{aligned}
\end{equation}

Hence, we can get
\begin{equation} \label{eq1111}
\begin{aligned}
G(\theta^*)- G(v^{t+1}_{[n]}) \leq [1-2H'' \beta(1+\frac{H''\beta}{2})] [G(\theta^*)- G(v^{t}_{[n]})]
\end{aligned}
\end{equation}
Here we denot $\xi =1-2H'' \beta(1+\frac{H''\beta}{2})$. 
That is to say,
\begin{equation} 
G(\theta^*)- G(v^{t+1}_{[n]}) \leq \xi [G(\theta^*)- G(v^{t}_{[n]})].
\end{equation}
Iteratively, we can have
\begin{equation} 
\begin{aligned}
G(\theta^*) - G(v^{NT_0}_{[N]}) &\leq \xi^{T_0} [G(\theta^*) - G(v^{(n-1)T_0}_{[n]})] \\
&= \xi^{T_0} [G(\theta^*)-G(v^{(n-1)T_0}_{[n-1]})] +\\&\xi^{T_0} [G(v^{(n-1)T_0}_{[n-1]}) - G(v^{(n-1)T_0}_{[n]})].
\end{aligned}
\end{equation}

\begin{table*}[tb]
	\centering
	\renewcommand\arraystretch{1.2}
	\setlength{\abovecaptionskip}{0.cm}
	\caption{Parameters of channel dataset}
	\begin{tabular}{cccccc}
		\hline
		Parameters &SIM-P &SIM-B&NOF&NCS &CWR \\
		\hline
		Environment	 &Rayleigh & Default  &	Fjord	& Shelf  &Reservoir \\	
		Range &-	&500m$\sim$8000m &	750m 	&540m  &1100m, 2100m, 6000m  \\		
		
		Water depth  &-&100m &	10m&	80m & 50m \\		
		Transmitter depl &- &Suspended &	Bottom&	Bottom & Suspended \\
		Receiver depl &- &Suspended &	Bottom&	Bottom & Suspended \\
		Doppler coverage &30Hz&Uncalculated&	7.8Hz&	31.4Hz &Uncalculated\\		
		
		
		\hline  
	\end{tabular}\\
\end{table*}
\begin{figure*}
	\centering
	
	
	\subfigure[SIM-B]{
		\begin{minipage}[b]{0.2\textwidth}
			\includegraphics[width=\textwidth]{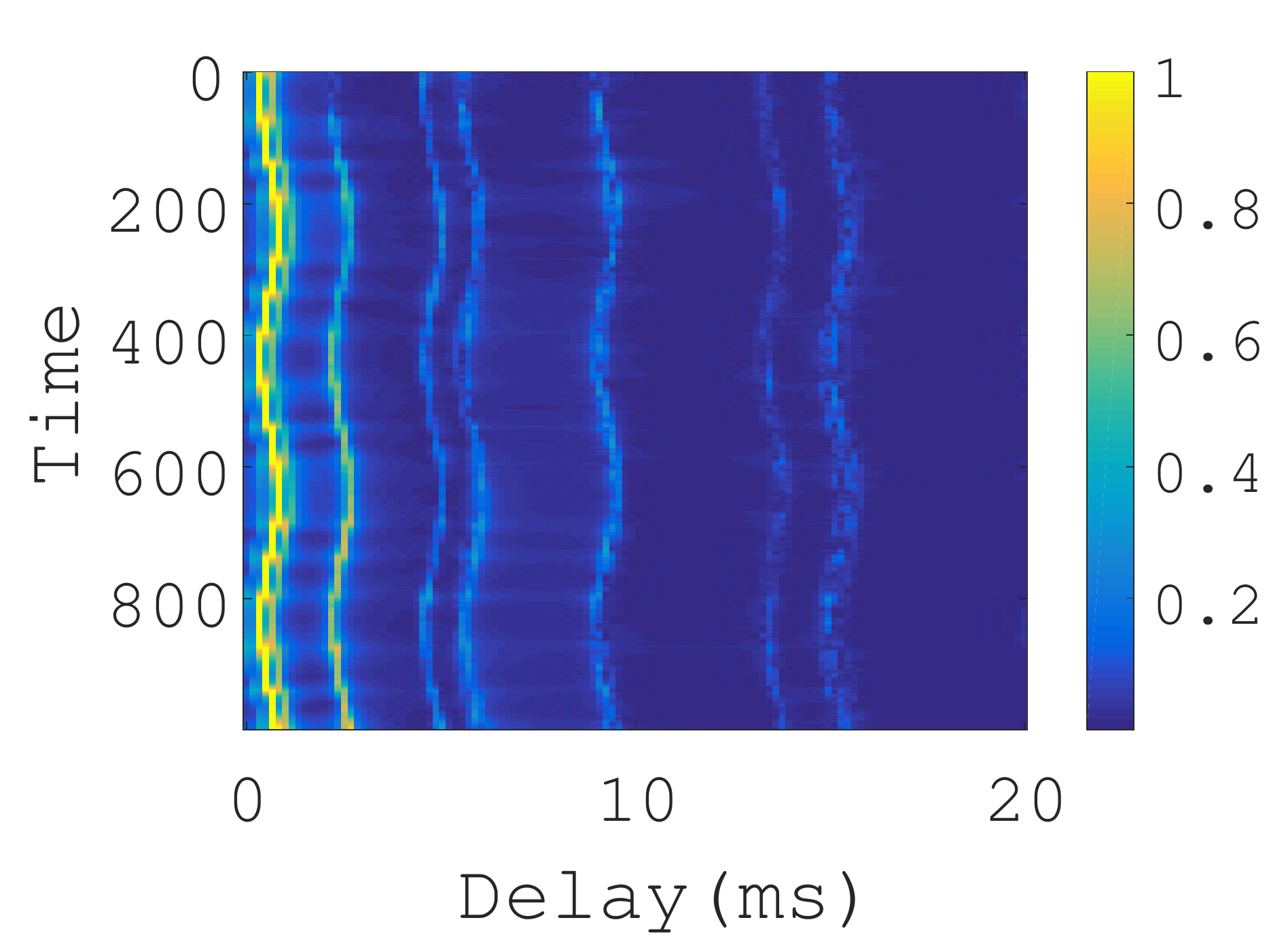}\label{09}
		\end{minipage}
	}
	\subfigure[NOF]{
		\begin{minipage}[b]{0.2\textwidth}
			\includegraphics[width=\textwidth]{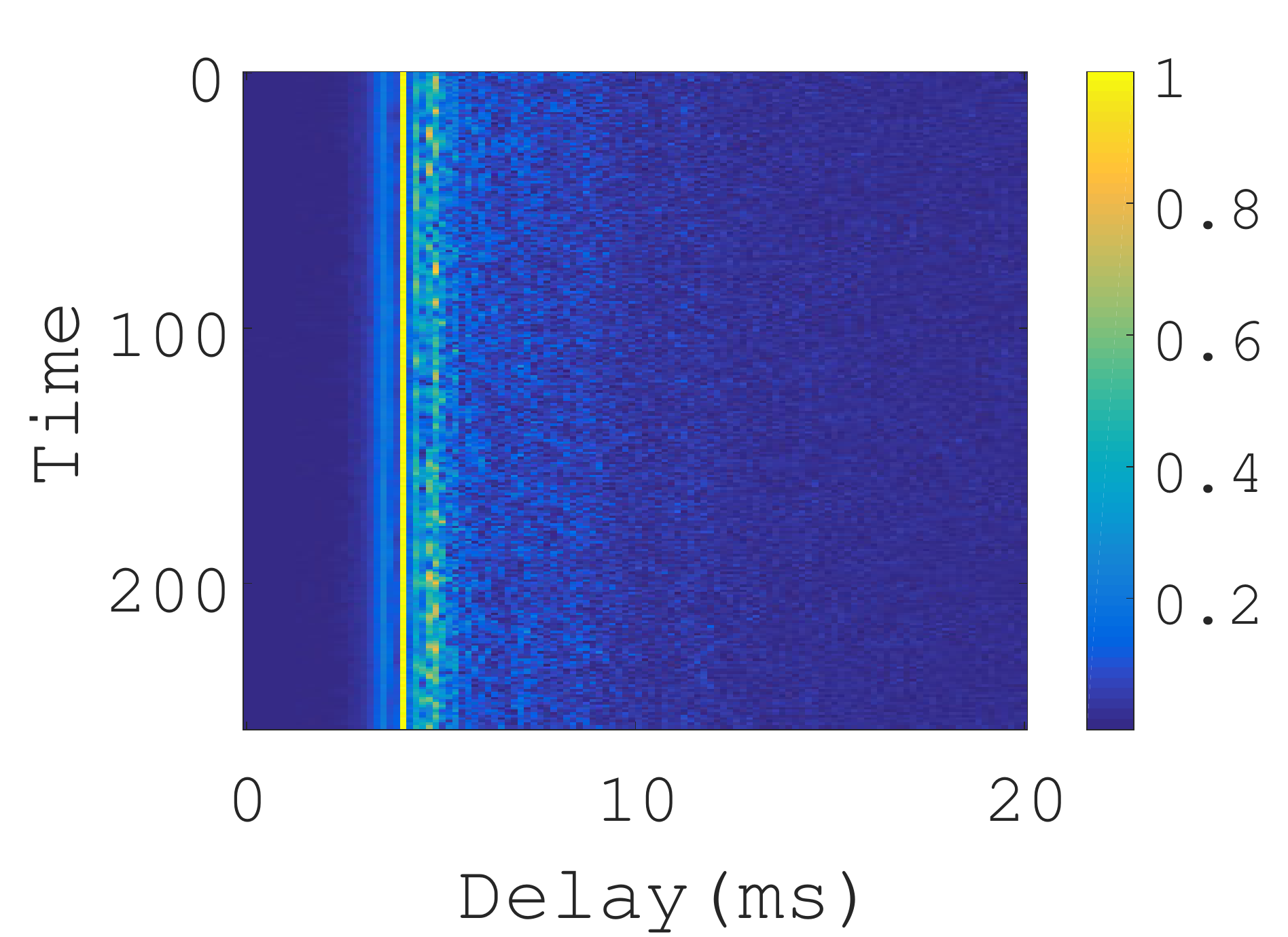}\label{075}
		\end{minipage}
	}
	\subfigure[NCS]{
		\begin{minipage}[b]{0.2\textwidth}
			\includegraphics[width=\textwidth]{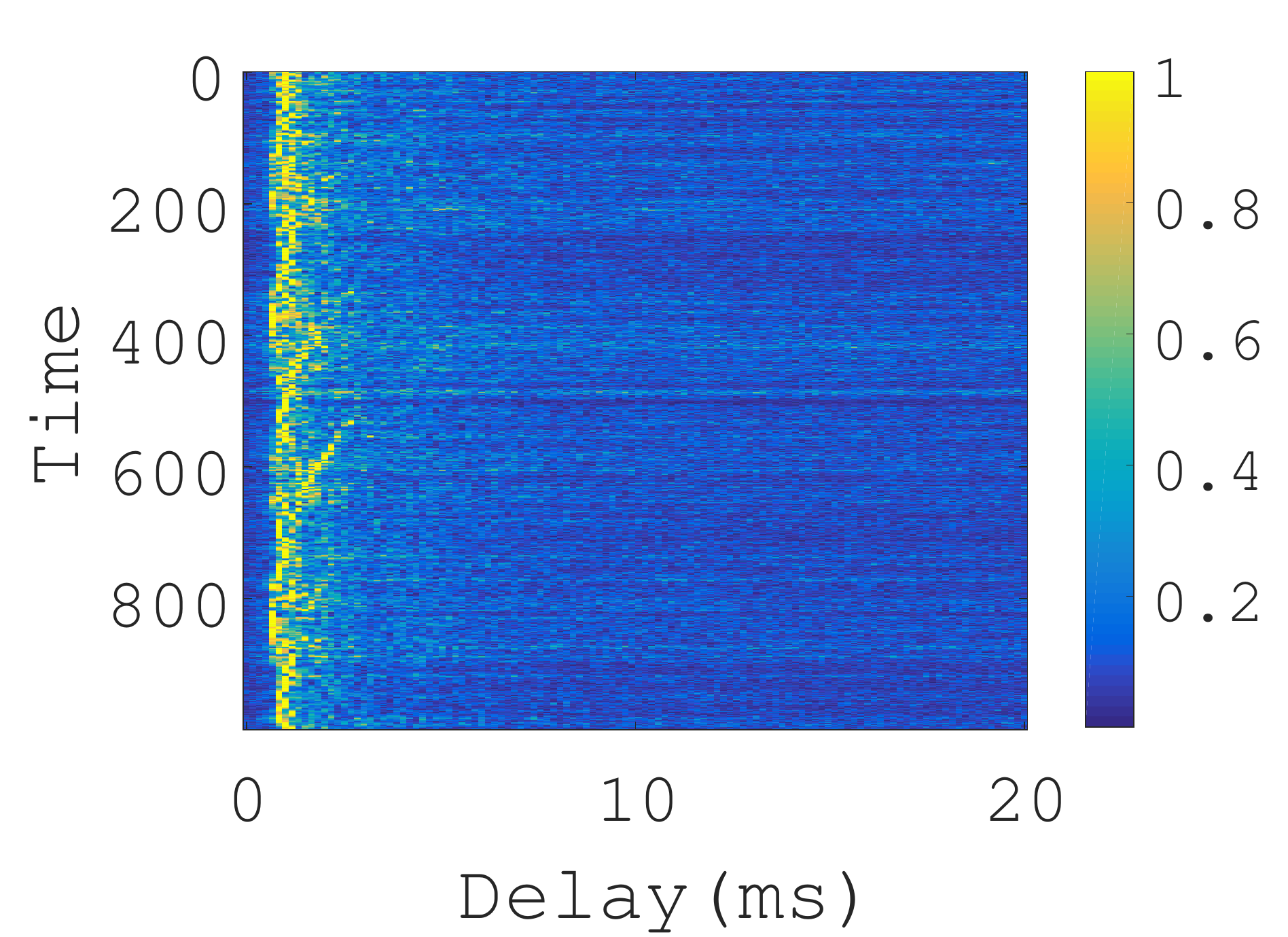}\label{08}
		\end{minipage}
	}	
	\subfigure[CWR]{
		\begin{minipage}[b]{0.2\textwidth}
			\includegraphics[width=\textwidth]{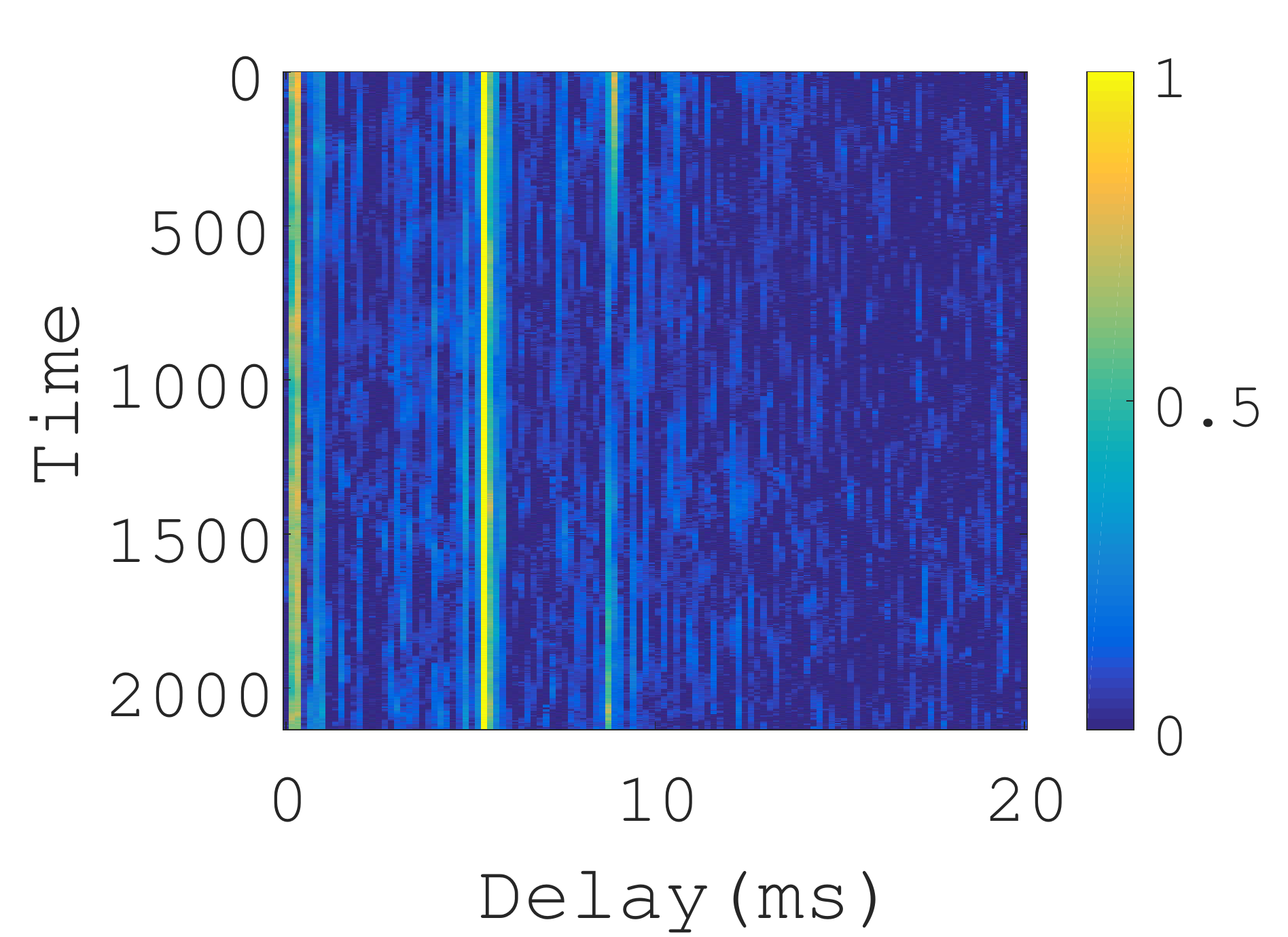}\label{Acceleration_Factor1}
		\end{minipage}
	}
	\caption{\small A snapshot of CIR dataset.}\label{ch_all}
\end{figure*}

Because$G(.)$ is $\mu''$-strong convex, we can get that $||G(\theta) -G(\theta') || \geq \mu''||\theta -\theta'||$.
Hence, the lower bound  $G(v^{(n-1)T_0}_{[n]})-G(v^{(n-1)T_0}_{[n-1]})$ is shown as follow
\begin{equation}
\begin{aligned}
&\mathbb{E}[G(v^{(n-1)T_0}_{[n]})-G(v^{(n-1)T_0}_{[n-1]})]\\
&=\mathbb{E}[G(\theta^{(n-1)T_0})-G(v^{(n-1)T_0}_{[n-1]})]\\
&\geq \mathbb{E}[\mu''||\theta^{(n-1)T_0} - v^{(n-1)T_0}_{[n-1]} ||]\\
&\geq  \mu'' m(T_0).
\end{aligned}
\end{equation}	
Iteratively and incrementally, we can get 
\begin{equation} 
\begin{aligned}
&\mathbb{E}[G(\theta^*) - G(v^{NT_0}_{[N]})] \\
&\leq  \xi^{T_0}\mathbb{E} [G(\theta^*)-G(v^{(n-1)T_0}_{[n-1]})] -\xi^{T_0} \mu'' m(T_0)\\
& \leq \xi^{(n-1)T_0}\mathbb{E} [G(\theta^*)-G(v^{0}_{[1]})] -\sum_{j=1}^{N-1}\xi^{jT_0} \mu'' m(T_0)
\end{aligned}
\end{equation}

After receiving updates from $T$-th rounds, the convergence gap of objective function can be expressed as 
\begin{equation} \label{gap}
\begin{aligned}
&\mathbb{E}[G(\theta^*) - G(\theta^{T})] \\
&= \mathbb{E}[G(\theta^*) - G(v_{[N]}^{Nt})]+\mathbb{E}[G(v_{[N]}^{Nt}) - G(v_{[N+1]}^{Nt})]\\
&\leq \xi ^T\mathbb{E}[G(\theta^*) - G(\theta^0)] - \sum_{j=1}^{N-1}\xi^{jT_0}  \mu'' m(T_0) - \mu'' m(T_0)\\
&\leq \xi^T\mathbb{E}[G(\theta^*) - G(\theta^0)]- \frac{ \mu'' m(T_0)}{1-\xi^{T_0}}
,
\end{aligned}
\end{equation}
The upper bound  of is  $\epsilon$ and  we have   $\mathbb{E}[G(\theta^0)-G(\theta^*)] \le n$ \cite{COCOA}. Thereby we complete the proof.


\section{ Dataset}
In addition, massive data is critical for deep learning. In radio frequency (RF) communication system,  the required data sets can be found online, such as DeepSig dataset\cite{deepsig},  RF channel dataset \cite{qin2019deep}.  But open source underwater acoustic communication dataset for learning algorithm is still blank. What's more, acoustic channel models and open-source software are foreseen as some of the key elements in the next generation of UWA COMMs research practices\cite{SongEditorial}.
The approximate channel models  can be implemented by the tapped delay line. We assume the signal is
bandlimited within bandwidth B  which can be described by discrete samples. Hence, formula (\ref{channelintergral}) can be expressed as
\begin{equation}\label{discreteCH}
r[k] = \sum_{k=0}^{K}g_k(t)x(t-kT_s)
\end{equation}
where $T_s$ is the sampling interval and $T_s=\frac{1}{B}$. The tap gain can be calculated by
\begin{equation}\label{eq-III.1}
g_k(t)=\int_{-\infty}^{+\infty}h(\tau,t) sinc(\frac{\tau-kT_s}{T_s})d\tau.
\end{equation}
where $sinc$ is sample function. Formula (\ref{discreteCH}) shows that $y(t)$ can be generated by passing $x(t)$
through a tapped delay line or FIR-filter with taps spaced $T_s$.
Next, we use three methods to get $h(\tau,t)$ for our dataset.
\subsection{Channel Impulse Response}
\subsubsection{Probability Model}Deep learning requires massive underwater acoustic channel impulse response to trained our net.
There are many ways to simulate underwater acoustic channel impulse response.
For generate massive CIR for learning, the probability model is selected to model and analyze the underwater acoustic channel. In this article,  we summarize the  experience of our predecessors, built a  simulation underwater acoustic Rayleigh channel dataset for shallow water horizontal communication. The UWA  multipath distribution can be assumed as a Rayleigh-distributed\cite{UWAOFDMbook}.  In this model,  maximum excess delay is set to 12ms, exponentially decaying is designed and the power attenuation coefficient is 0.66 dB per tap\cite{ebihara2015doppler}. And the UWA channel Doppler spread is considered by a bell-shaped function with $a$  equals 9 as the formula (12). 
\begin{equation}\label{eq-III.1}
S(f) = \frac{{\sqrt a }}{{\pi {f_d}\left\{ {1 + a\left( {\frac{f}{{{f_d}}}} \right)} \right\}}}  \left| f \right| \le {f_d}
\end{equation}

\subsubsection{Propagation Model}
BELLHOP (SIM-B) is a  widely-known UWA channel simulation method.  
Taking it into account, we create a part of data by \cite{Qarabaqi}. The horizontal distance between the receiving end and the transmitting end varies between 500m and 8000m, randomly.
The vertical distance varies from 10m to 90m, randomly.
\subsubsection{Measured CIR}
In this subsection, we provide a multi-scene validation  dataset, containing simulation and measured CIRs with multi-communication environments, and multi-communication ranges.
In addition, the CIRs measured under different environments are considered. 
The raw CIRs were measured at Norway-Oslofjord (NOF), Norway-Continental Shelf (NCS)\cite{van2017watermark} and China-Wanlu Reservoir (CWR). It is  worth noting that, in CWR, CIRs from different distances were collected. Both receiver and receiver terminals were located on two ships for long-distance communication test, with distances of 1100m, 2100m, and 6000m, respectively. 
After this, we consider two  specific condition as  data augmentation.

\subsection{Data Augmentation}

\subsubsection{Symbol Time Offset}If the synchronization error or loss of synchronization occurs in the communication system, the performance of the communication system will be reduced or the communication failure will occur. In order to ensure that the system can reliably detect the synchronous signal, it is more difficult to detect the synchronous signal of underwater communication system due to the complexity of the above-mentioned conditions of underwater acoustic signal transmission


\begin{equation}\label{eq-III.1}
\hat r'(t) = r(t+\delta),
\end{equation} 
where $\delta$ means the  deviation caused by inaccuracy synchronization.
\subsubsection{Doppler Shift}
In UWA channel, the effect of platform
motion on a wideband signal is more accurately modeled
as a complete time scaling (expansion or compression) of
the signal waveform. We assume that the mean Doppler shift can not be removed from the sounding data before the correlation.
The signal can be expressed as
\begin{equation}\label{eq-III.1}
r(t) = s((1+\alpha)t),
\end{equation} 
where $s(t)$ and $r(t)$ are the source and Doppler-shifted
received signals, respectively. The relative Doppler shift $\alpha$
is defined as the ration of the relative platform speed to
the sound speed, which can be calculated by
\begin{equation}\label{eq-III.1}
\alpha  =\frac{\delta f}{f_c} =  \frac{v}{c},
\end{equation} 
where $c$ means the speed of sound in the water and  $v$ denotes the  relative speed of the transmitter and receiver.
\section{Simulation Results}
\begin{figure}[t] 
	\centering
	\includegraphics[width=8cm]{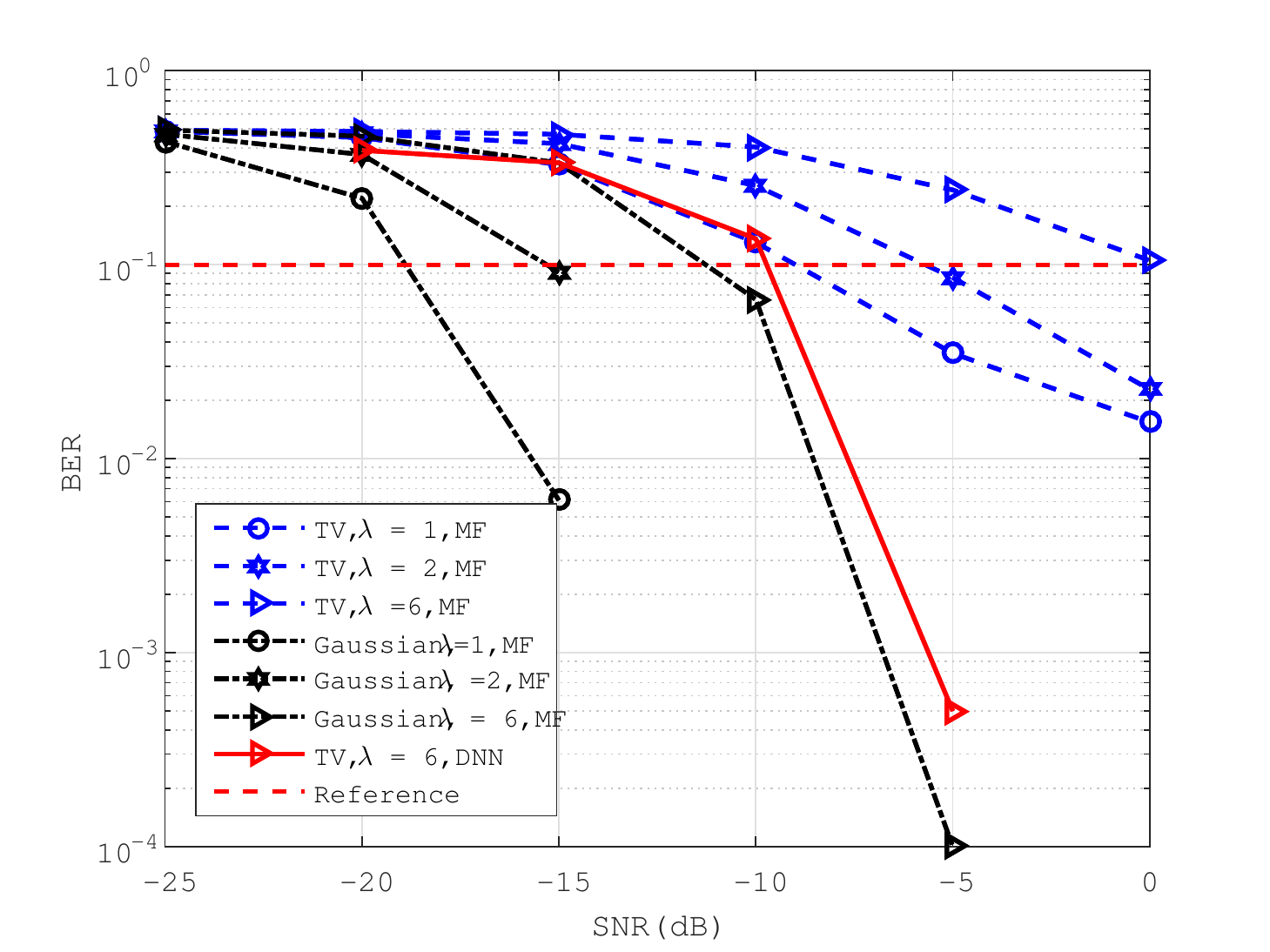}\\
	\caption{ BER curves of DNN and MF under different downsample factor.}\label{SAMPLE}
\end{figure}
\begin{figure}[t] 
	\centering
	\includegraphics[width=8cm]{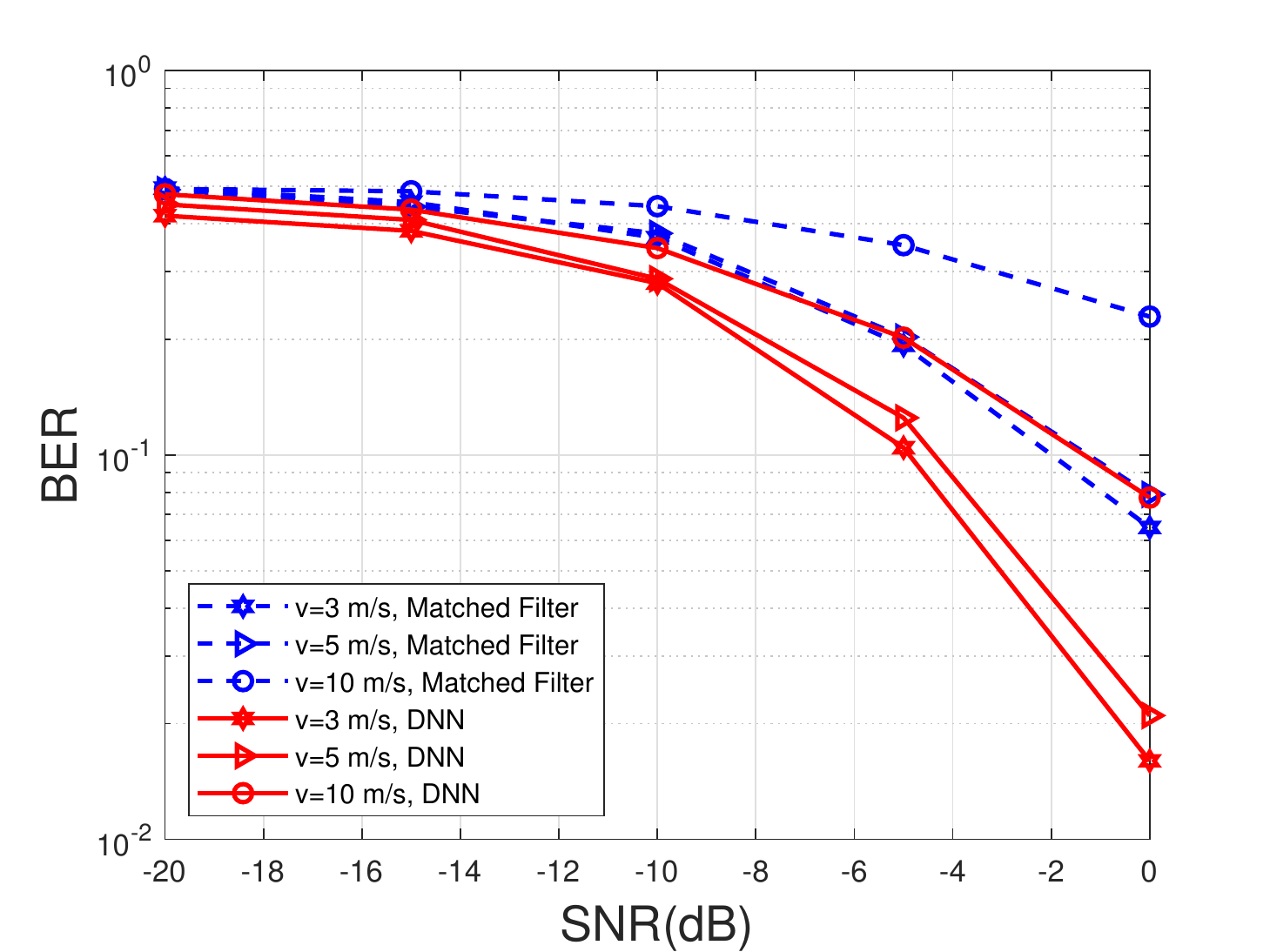}\\
	\caption{ BER curves of DNN and MF ($\lambda$=1) under different relative speed.}\label{DOPPLER}
\end{figure}

\begin{figure}[t] 
	\centering
	\includegraphics[width=8cm]{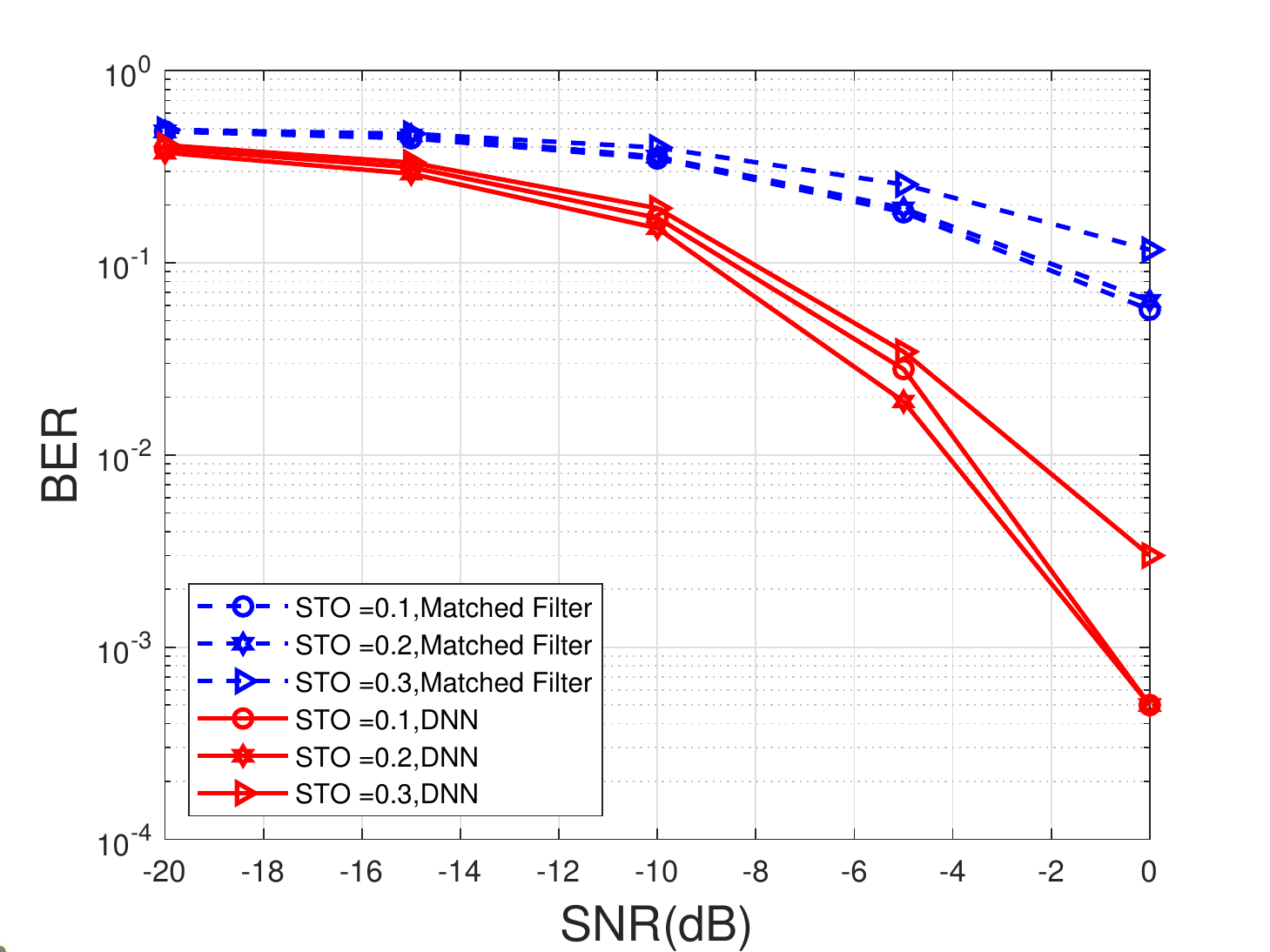}\\
	\caption{ BER curves of DNN and MF ($\lambda$=1)  under different STO.}\label{STO}
\end{figure}

\begin{figure}[t] 
	\centering
	\includegraphics[width=8cm]{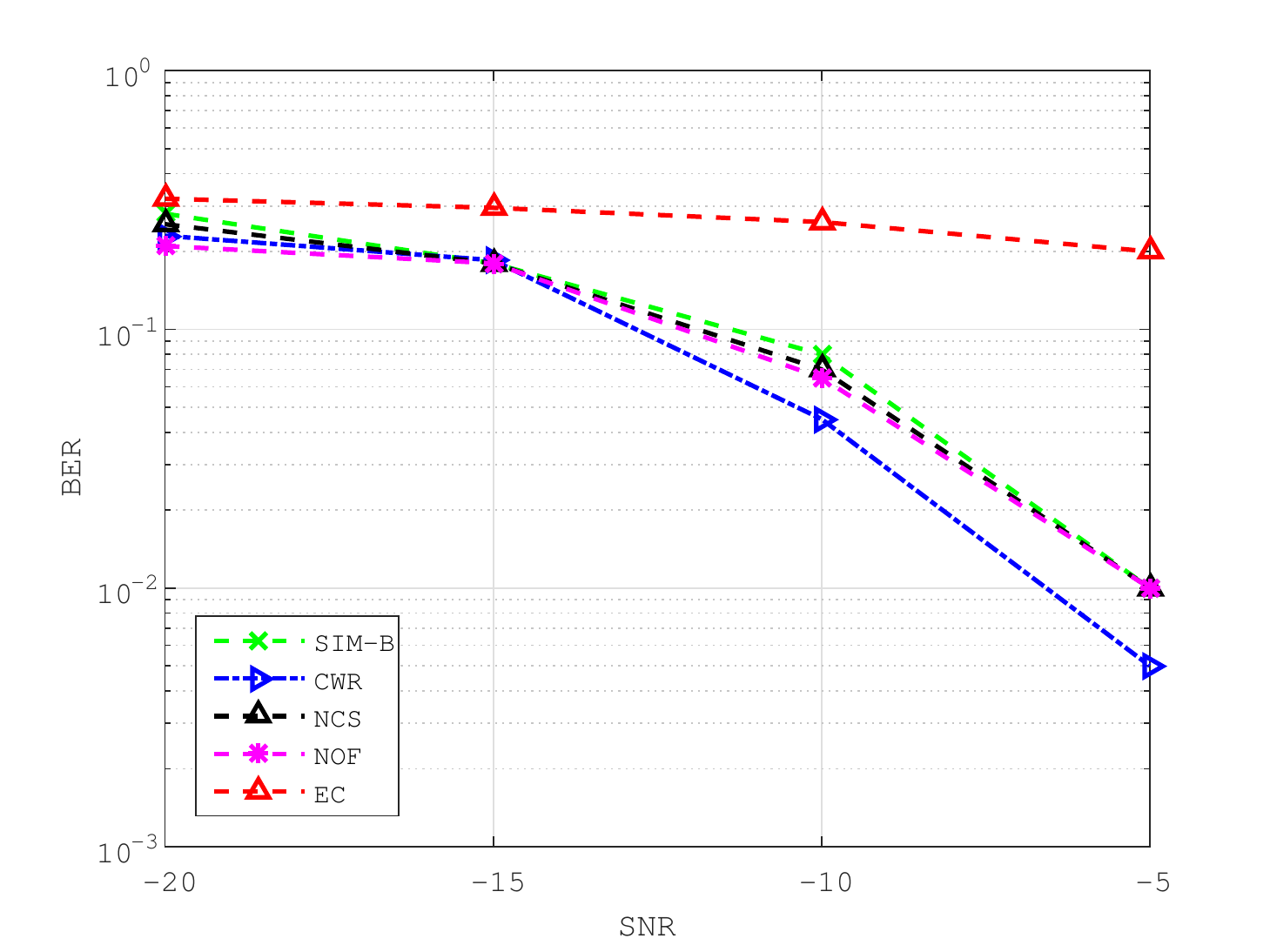}\\
	\caption{ C-DNN receiver generalization under different conditions.}\label{generalization}
\end{figure}
In this section, in order to evaluate  the  effect of proposed system.  The results are divided into two parts. One is focus on amazing performance of DL based chirp receiver. The other is the convergency performance of FML enhanced communications. Specially, we first introduce the parameters we used in our simulations. Then, we start with  the  performance of single node  to illustrate how great deep learning is for physical layer, especially for complex UWA COMMs.
\subsection{Parameters}
In our experiments, Pytorch and Matlab are used as development framework.  Each chirp frame contains  200 symbols and  each symbol duration is 10ms.  The input contains real parts. Every symbol  is predicted independently. The parameters of  receiver DNN scheme are shown in table I.
In the following experiment, if there is no special explanation, the local adaptation rate $\alpha$ is  0.001 and the update rate $\beta$ is 0.0001.  The parameters the avaliable ratio, the number of access users and local epoch are configured with $G=0.3$, $N=10$ and $T=T_0$.
For  buoy node, each local dataset contains 1000 symbols for training.

\renewcommand{\algorithmicrequire}{ \textbf{Input:}} 
\renewcommand{\algorithmicensure}{ \textbf{Output:}} 

\subsection{Performance of Single-node}

\subsubsection{Under different down-sampling factors}

Figure.\ref{SAMPLE} shows BER curves of DNN and MF under different downsample factor and different channel conditions.
From the results, we can discover a distinct characteristic that  the smaller sample factor we get, the lower bit error rate can be obtained. 
This is due to the smaller sample factor means the more sample points which influence  signal-to-noise ratio of the used signal in detection.  
Besides, we can find that the C-DNN receiver can reach the almost ideal bit error rate performance. 
\subsubsection{Under different STO and DOP}

The accuracy vs SNR performance of C-DNN and matched filter are presented in Fig.4. Through observation, we can see 
For deep learning based receiver, not only BER is the key characteristic to decipt communication system performance, but also needs the comparison of train loss and the valid loss to reflect the generalization capability of the receiver. 
Fig.3 The DNN receiver BER performance can be 
There is a fact that the receiver  is easy to overfit as shown in Fig.4. When receiver is trained online, the loss of model on  training set is better than that on validate set.   Even so, when the receiver is deployed offline, the receiver performance can better than traditional  minimum meansquare error algorithm based receiver . If the generalization ability of the model can be further improved, the receiver performance will be further improved. Deep learning based receiver  has great potential.

\subsubsection{Complexity Analysis}
From Table \ref{compareC}, it is obvious that, DL-receiver has lower  complexity.  
The computations of matched filter are all from correlation calculation. 
The total number of additions (ADD) used by  matched filter is $N_1+N_2-1$
and multiplications (MUL) is $N_1(N_1+N_2-1)$, where $N_1=N_2= \frac{Tf_s}{\lambda}$.
The total number of additions used by  matched filter is 
$C^{MF}_{ADD}=2N_1-1,$
The total number of multiplications is 
$C^{MF}_{MUL}=2N_1^2-2N_1,$
The total number of computations are presented in terms of the elementary operations is $C^{MF}_{TOTAL}= C^{MF}_{ADD}+C^{MF}_{MUL}=2N_1^2-1$.
The total number of additions used by  DNN is
$C^{DNN}_{ADD}=\sum_{l=1}^{L}N_l =\frac{15}{8}N_1+1,$ 
The total number of multiplications is 
$C^{DNN}_{MUL}=N_1^2+\sum_{l=2}^{L}N_l N_{l-1}= \frac{53}{32}N_1^2+\frac{1}{8}N_1,$ 
The total number of non-linear activations,
amount to $C^{DNN}_{NAV}=\sum_{l=1}^{L}N_l= \frac{15}{8}N_1+1.$
The total number of computations are presented in terms of the elementary operations is 
$C^{DNN}_{TOTAL}= C^{DNN}_{ADD}+C^{DNN}_{MUL}+C^{DNN}_{NAV} =\frac{53}{32}N_1^2+\frac{31}{8}N_1+2.$
It can be concluded from the above, the $C^{DNN}_{TOTAL} \le C^{MF}_{TOTAL} $, because  $N_1$  always is a large number.

\begin{table}[b]
	\centering
	\renewcommand\arraystretch{1.2}
	\renewcommand\tabcolsep{3.0pt}
	\setlength{\abovecaptionskip}{0.cm}
	\caption{Complexity analysis for DNN and MF.}
	\begin{tabular}{ccccc}
		\hline
		Operations&MF ($\lambda$ =1)&	MF ($\lambda$  =2)&	MF ($\lambda$  =6)	&DNN ($\lambda$  =6)\\
		\hline
		ADD & 1,919	&	959&	319&	301\\
		MUL	&	1,842,240&460,320&51,040&	42,420\\
		Non-Linear Activations &	-&	-&	-&	301\\
		Total  &	1,844,159&	461,279&	51,359&	43,022\\
		DNN Advantage  & 4186.5\%&		972.2\%&	19.4\%&	-\\
		
		\hline  
	\end{tabular}\\\label{compareC}
\end{table}

\subsection{Generalization of C-DNN}
 Figure.\ref{generalization} shows that C-DNN can achieve similar performance under channels that have never appeared in the training dataset. 
We use simulation data to train our C-DNN and test its performance under simulated and measured dataset. We test C-DNN under  different scenarios, where NCS, NOF and CWR  have diverse communication environments. It is worth mentioning that a variety of distance are employed in CWR. Hence, we can  believe  the C-DNN receiver can handle multiple scenarios with only a small loss of accuracy.  That is to say, C-DNN is robust can deal with different channals. Moreover, we also test C-DNN performance under emergency conditions (EC) that the data augmentation cannot cover all Doppler shift and symbol timing offset cases. Unfortunately, under EC condition, the C-DNN  encounter performance degradation because the DL-model meet something it had never seen before. 

\begin{figure}[t] 
	\centering
	\includegraphics[width=8cm]{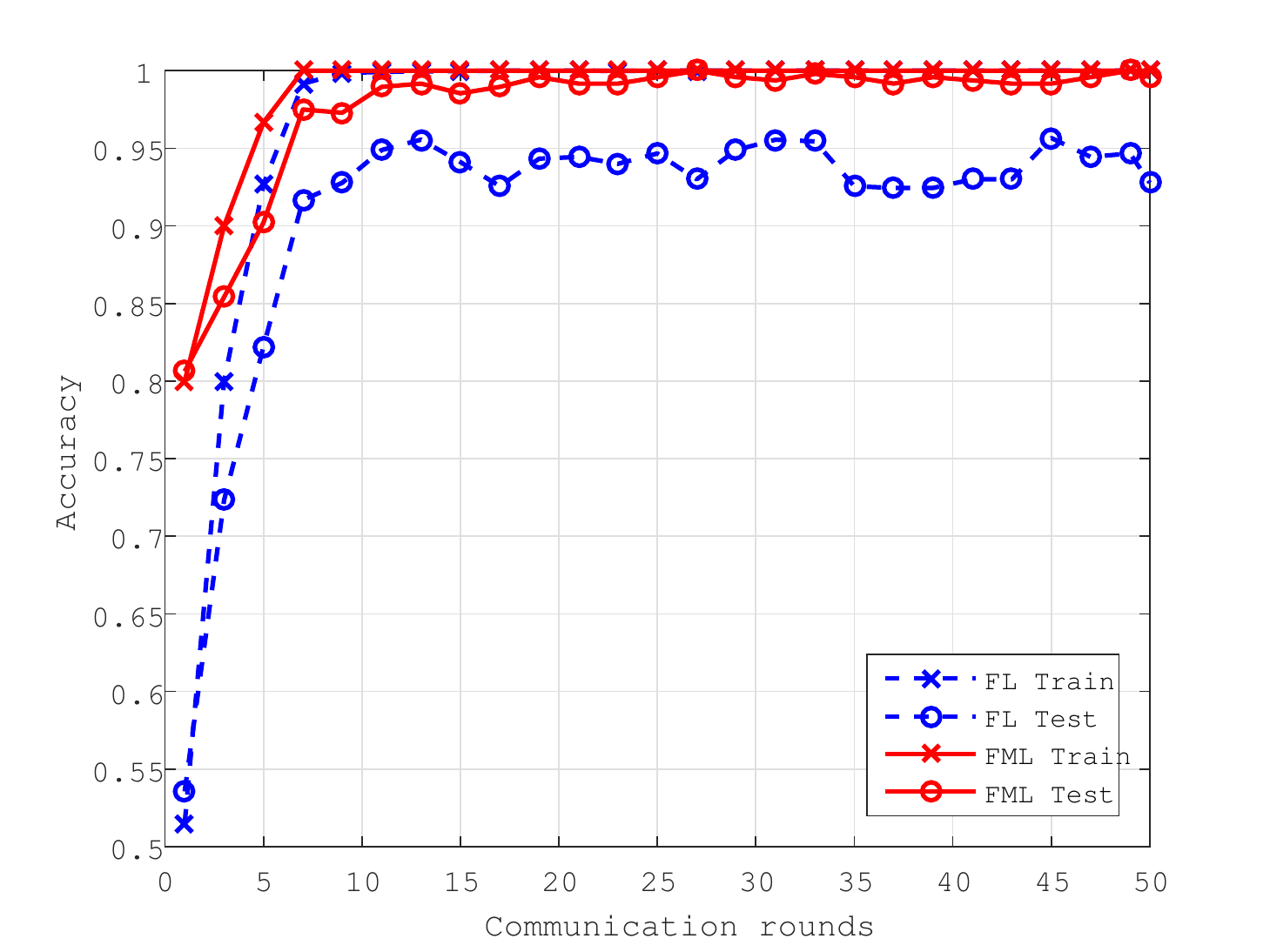}\\
	\caption{ Accuracy  vs communication rounds  peformance of federated learning and federated meta learning.}\label{FLvsFML}
\end{figure}

\subsection{Performance of FML}
\subsubsection{With federated learning} From Fig.\ref{FLvsFML}, we can find that FL and FML have the similar convergence performance under training dataset. But for test stage, test dataset considered, FL accuracy will decrease because the insufficient generalization of the model. FML utilize the then fine-tune the network with the labeled data in the target dataset.  In genearl, if the source dataset and target dataset are highly related, a FL algorithm would perform well, without the need for fine-tuning the DL-receivers according to the target environment, which owes to generalization of it. However, most of the time, it's hard to make source and target dataset equally distributed. Hence, FML has wider application scenarios. 
\subsubsection{With different local epochs}
\begin{figure}[t] 
	\centering
	\includegraphics[width=8cm]{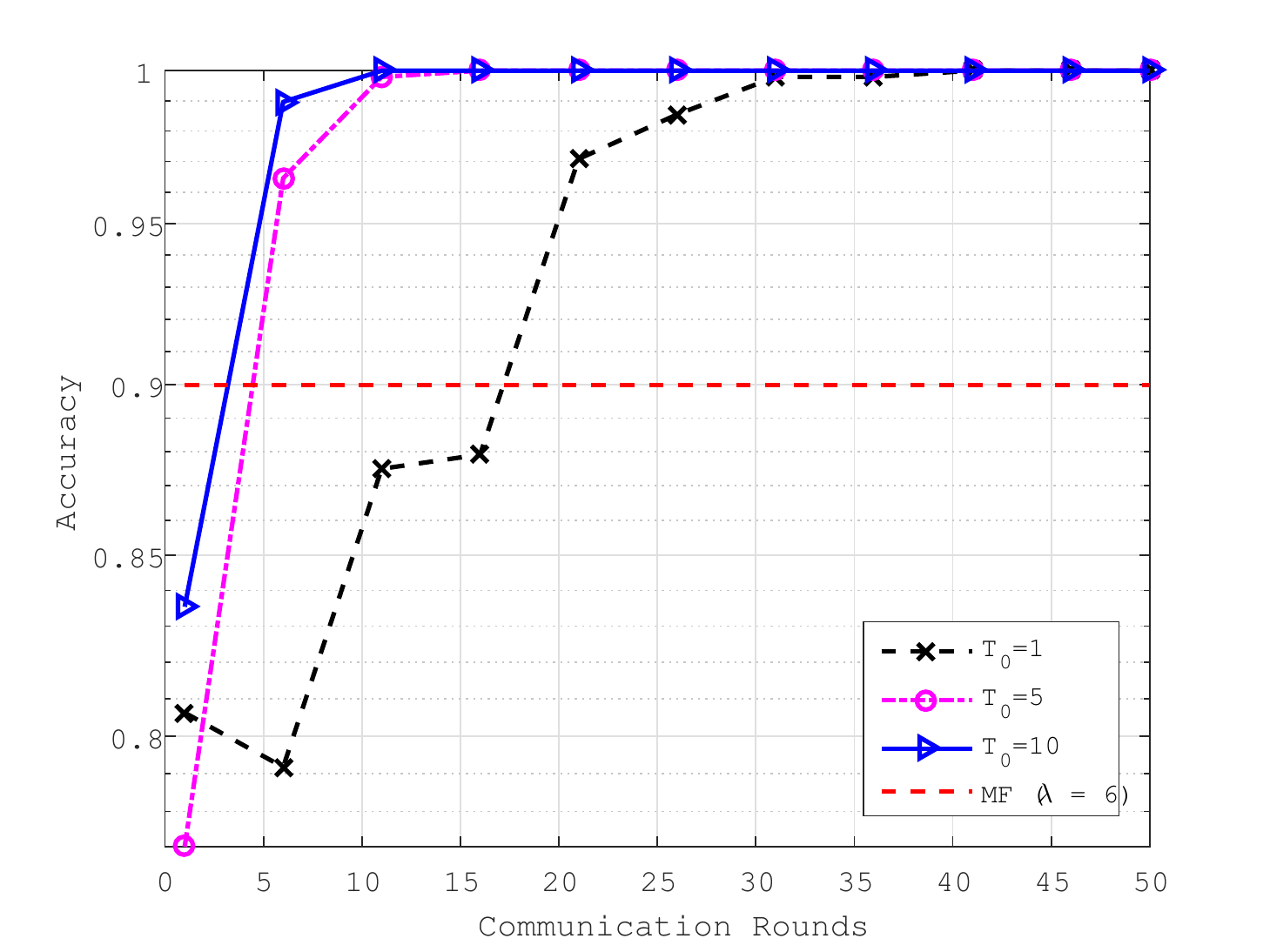}\\
	\caption{ Accuracy  vs communication rounds with different local epoch.}\label{localepoch}
\end{figure}

Performance of accuracy vs communication rounds with different local epoch is shown in  Fig.\ref{localepoch}. From the results we can find that as $T_0$ increase, so does the accuracy of DL-receiver. The case with local epoch $T=5T_0$ converges  more than $10$ communication rounds faster than that with $T=T_0$. However, as the number of local training epoch increases, this advantage will decrease. The case with local epoch $T=5T_0$ and that with  $T=10T_0$  have the similar convergence rate. 
Formula (\ref{impans}) tell us that $K$ and $m(T)$ are increments of $T_0$. Hence, the simulation results verify the Theorem 2 that the convergency gap increase with  $T_0$ under a  fixed communication rounds $T$.  This result can guide us to balance the consume of communication rounds and local computations in the real system.

\subsubsection{With different data volumes}
Figure.\ref{differentVolumes} shows  the variation of accuracy compared  and communication rounds under different data volumes on a single node. From the results, we can find that with the number of users increasing,  the convergence of the whole system becomes faster. 
It is easy to see that  the larger  data volume used for training, the better system performance can be got. But in a real system we can't overdo the amount of local data, because of the  distributed data storage, especially in the ocean. Most importantly, we compare three kind of data volume to understand the influence of data volume to single node.

\begin{figure}[t] 
	\centering
	\includegraphics[width=8cm]{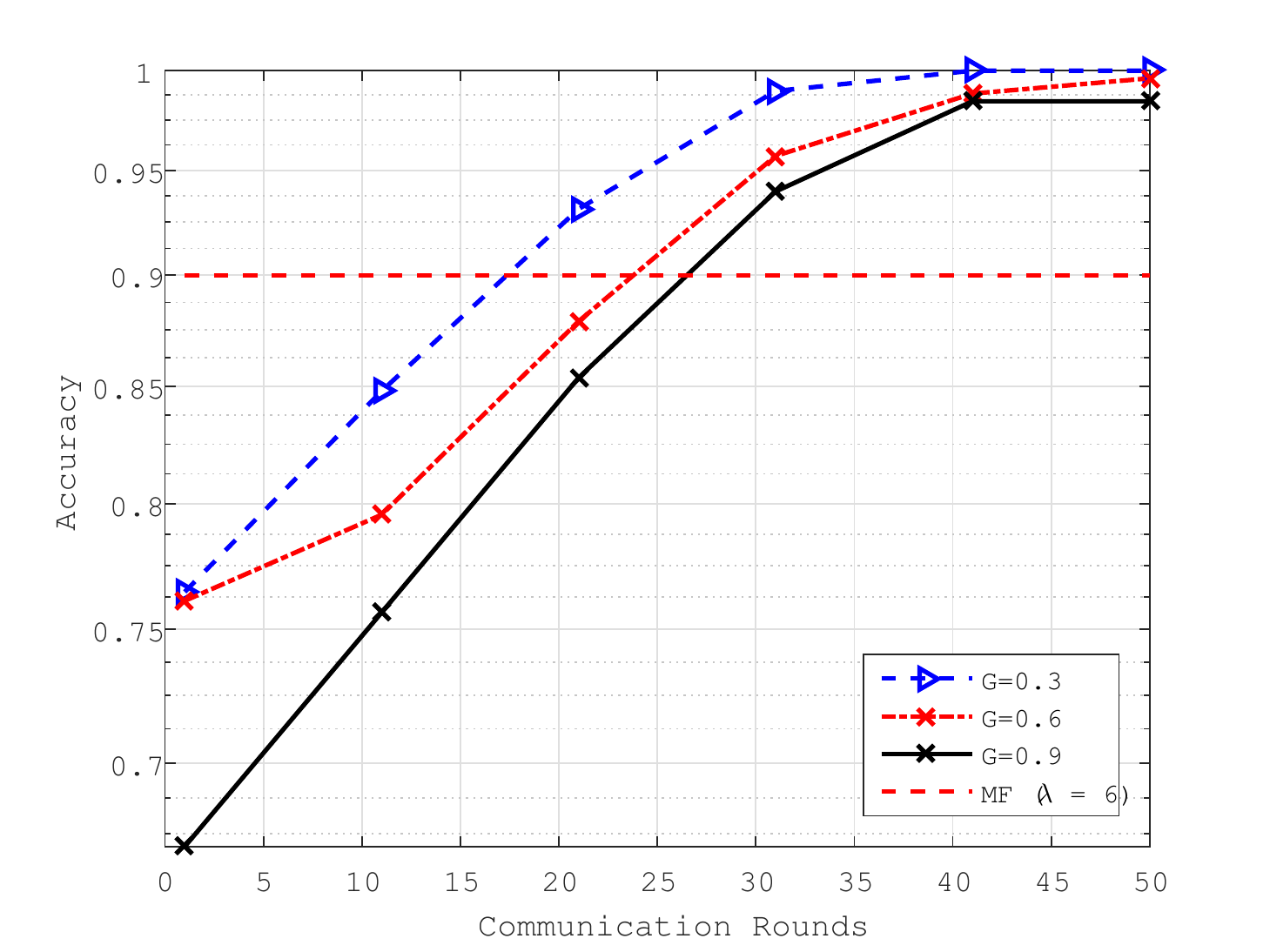}\\
	\caption{  Accuracy  vs communication rounds with different available channel ratio $G$.}\label{avaliableG}
\end{figure}
\begin{figure}[t] 
	\centering
	\includegraphics[width=8cm]{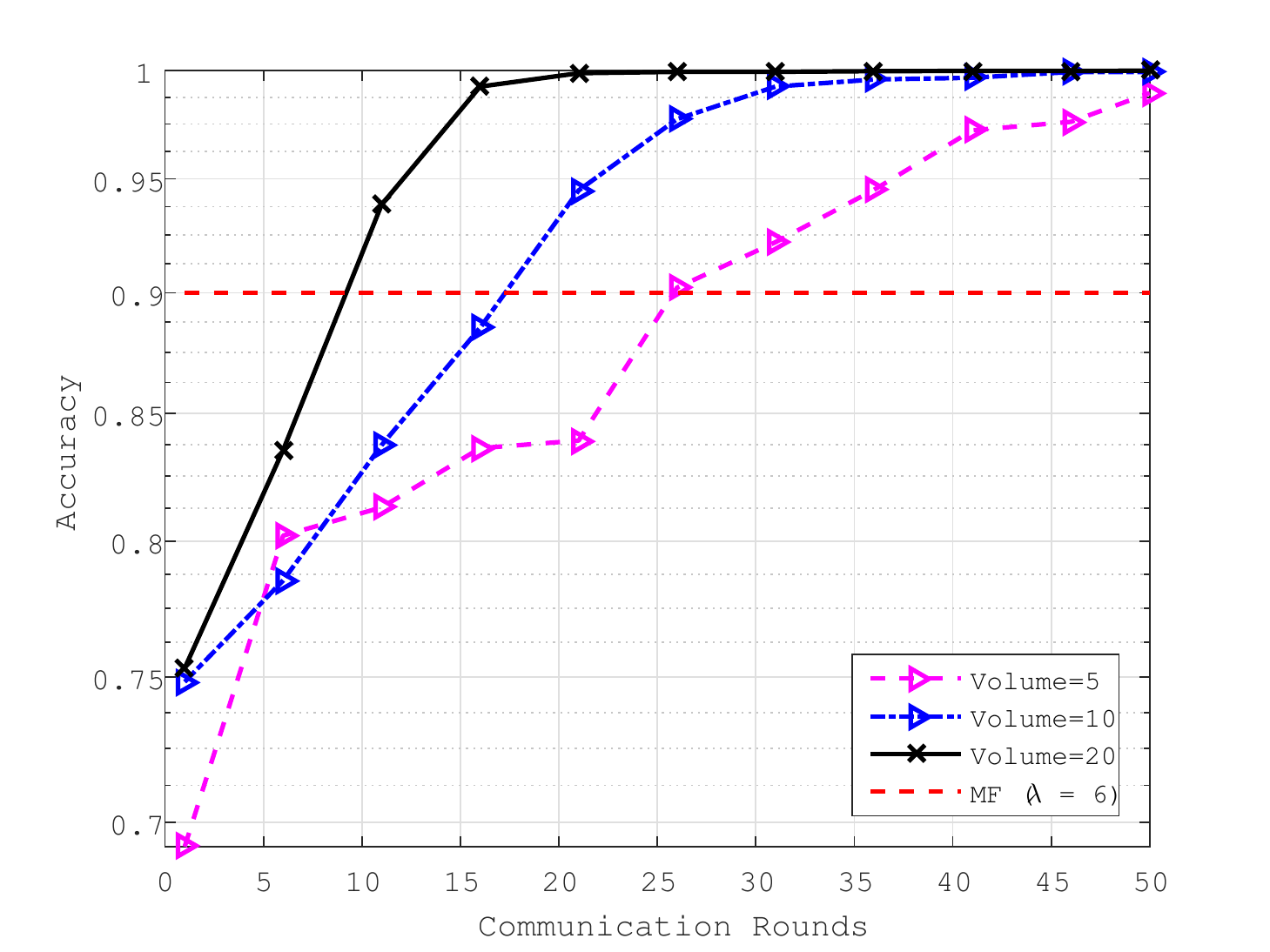}\\
	\caption{ Accuracy  vs communication rounds with different data volumes on a single node.}\label{differentVolumes}
\end{figure}

\begin{figure}[t] 
	\centering
	\includegraphics[width=8cm]{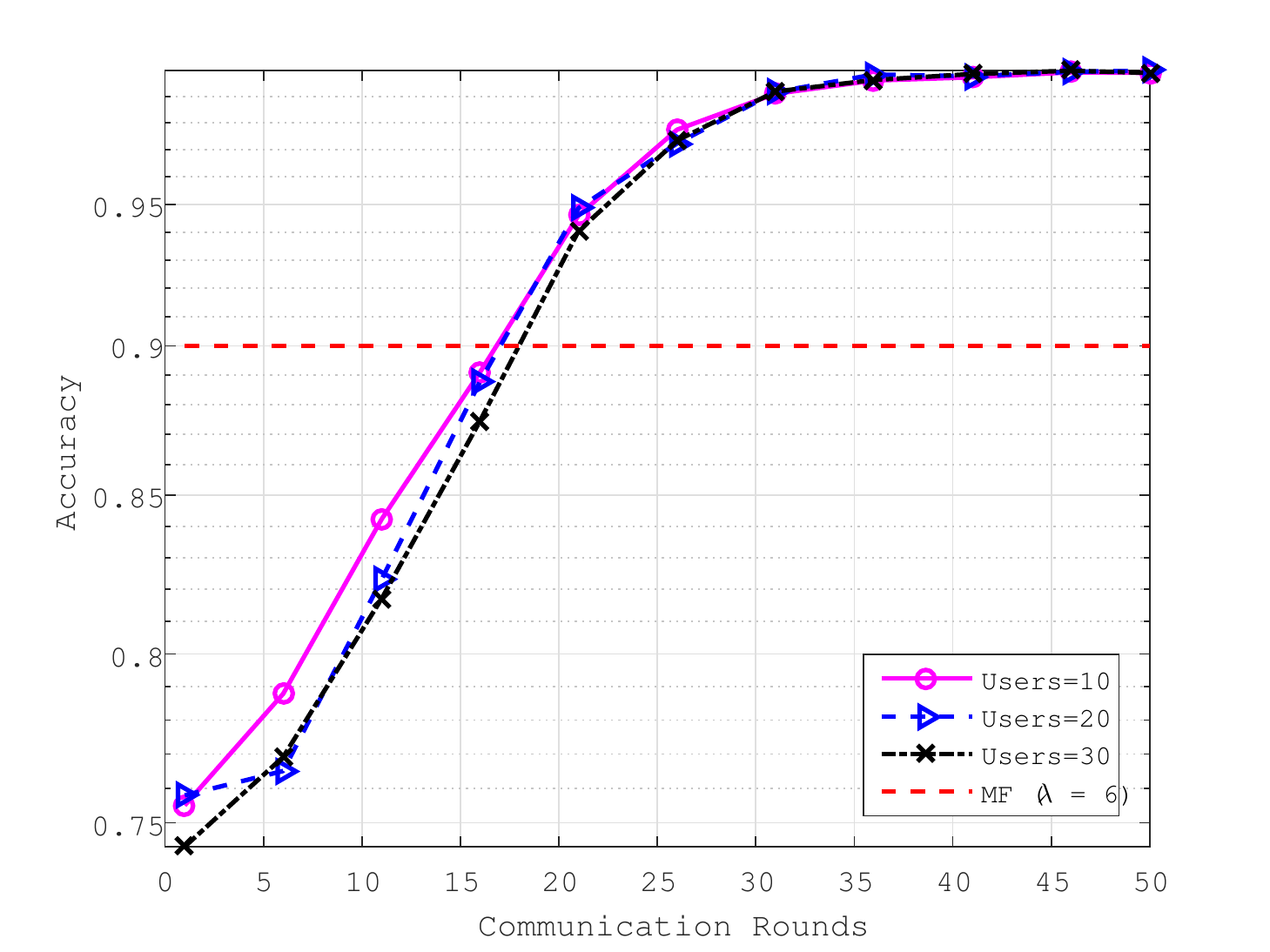}\\
	\caption{ Accuracy  vs communication rounds with different numbers of access users.}\label{differentUsers}
\end{figure}

\subsubsection{With different number of channels}
Figure.\ref{avaliableG} and Fig.\ref{differentUsers} indicate the  effect of the number of users and the impact of the number of channels available. We can draw the following conclusions. At a certain number of accessible channels, with the increasing of access users,
the higher accuracy of DL-receiver we can get.   Meanwhile,  at a certain number of access users, the higher  access rate, the slower the convergence rate. We can  explain it qualitatively and quantitatively. For quantitative analysis, with the increasing of users, the  convergence rate  $T_z$ will become slower. Formula (\ref{impans}) tell us $N$ and $G$ affect the convengence rate which is  inversely proportional relationship.  From qualitative aspect that as the number of users increases, leading to riching the differences  of training data, it slows down the convergence.


%

%
%

\section{Conclusions}\label{conclusions}
 Deep learning have great potential in underwater acoustic communication system but the disadvantage is sensitive to the distribution of training data.  Therefore, considering the generalization of DL-based applications, we utilize the  acoustic radio cooperation characteristic of OoT,  we proposed 
a federated Meta Learning Enhanced Acoustic Radio Cooperative Framework for Ocean of Things, which take advantage of the data distributed on surface nodes. Through this method, we can achieve transfer learning. 
We take  UWA chirp communications  as an example, which can provide stable UWA COMMs for Ocean of Things.
In order to overcome UWA doppler shift and symbol time offset, we proposed C-DNN receiver  based on data driven deep learning. Besides, to understand its performance, a comprehensive
convergence analysis framework for FML with  random schedule  in wireless is developed. This work represents the first
attempt to combine FML and DL in physical layer.
For future work, we will consider the framework of the
current work to sea trial. As another
interesting direction, the proposed design only for OoT device can be extended to the IoUT
scenario by using underwater unmanned submarine vehicle in an effort to utilizing  the distributed data,  further accelerating the learning process. 



%
%
%

\end{document}